\documentclass[structabstract]{aa}
\usepackage{amsmath}
\usepackage{graphicx}
\usepackage{aas_macros}
\usepackage{txfonts}
\usepackage{natbib}
\usepackage[english]{babel} 
\everymath{\displaystyle}

\def\pd#1#2{{\partial #1 \over \partial #2}}

\titlerunning{Modelling gamma-ray binary emission based on relativistic hydrodynamics}
\authorrunning{Dubus, Lamberts, Fromang}

\begin{document}
\date{Accepted . Received ; in original form \today}
	\title{Modelling the high-energy emission from gamma-ray binaries using numerical relativistic hydrodynamics}
	
	\author{G. Dubus\inst{1,2} \and A. Lamberts\inst{3}	\and	S. Fromang\inst{4}	}
	\institute{
	Univ. Grenoble Alpes, IPAG, F-38000 Grenoble, France
	\and
	CNRS, IPAG, F-38000 Grenoble, France
	\and
	Department of Physics, University of Wisconsin-Milwaukee, Milwaukee WI 53201, USA
	\and
	Laboratoire AIM, CEA/DSM--CNRS--Universit\'e Paris 7, Irfu/Service d'Astrophysique, CEA-Saclay, 91191 Gif-sur-Yvette, France	}

\begin{abstract}
{Detailed modeling of the high-energy emission from gamma-ray binaries has been propounded as a path to pulsar wind physics.}
{Fulfilling this ambition requires a coherent model of the flow and its emission in the region where the pulsar wind interacts with the stellar wind of its companion.}
{We developed a code that follows the evolution and emission of electrons in the shocked pulsar wind based on inputs from a relativistic hydrodynamical simulation. The code is used to model the well-documented spectral energy distribution and orbital modulations from LS 5039.}
{The pulsar wind is fully confined by a bow shock and a back shock. The particles are distributed into a narrow Maxwellian, emitting mostly GeV photons, and a power law radiating very efficiently over a broad energy range from X-rays to TeV gamma rays. Most of the emission arises from the apex of the bow shock. Doppler boosting shapes the X-ray and VHE lightcurves, constraining the system inclination to $i\approx 35\degr$. There is a tension between the hard VHE spectrum and the level of X-ray to MeV emission, which requires differing magnetic field intensities that are hard to achieve with a constant magnetisation $\sigma$ and Lorentz factor $\Gamma_{p}$ of the pulsar wind. Our best compromise implies $\sigma\approx 1$ and $\Gamma_{p}\approx 5\times 10^{3}$, respectively higher and lower than the typical values in pulsar wind nebulae.}
{The high value of $\sigma$ derived here, where the wind is confined close to the pulsar, supports the classical picture that has pulsar winds highly magnetised at launch. However, such magnetisations will require further investigations to be based on relativistic MHD simulations.}
\end{abstract}
\maketitle
\label{first page}

\section{Introduction}
Gamma-ray binaries are composed of a massive star in orbit with a compact object and characterized by dominant radiative output in the gamma-ray ($\ga$ MeV) range (see \citealt{2013A&ARv..21...64D} for a review). The compact object is widely thought to be a rotation-powered pulsar, although this remains to be proven for most systems. The interaction of the pulsar wind with the massive star wind ends up dissipating part of the pulsar's rotation power through particle acceleration. Gamma-ray binaries offer an opportunity to study the processes involved in pulsar wind nebulae on much smaller scales, notably how an $e^{+}e^{-}$ relativistic wind is launched from a rotating magnetosphere and how its energy is released at the termination shock.  

Observations of gamma-ray  binaries show flux variations tied to the orbital phase in most of them. Theoretical models have focused on relating these variations to changes in the line-of-sight and/or the conditions at the termination shock as the pulsar follows its eccentric orbit. Amongst the different systems, \object{LS 5039} constitutes a useful testbed because of its regular, well-documented modulations in X-rays \citep{Takahashi:2008vu}, low energy (LE, $1-100$ MeV, \citealt{2014A&A...565A..38C}), high energy (HE, $0.1-100\,$GeV, \citealt{2009ApJ...706L..56ABIS}) and very high energy (VHE , $\geq 100\,$GeV, \citealt{Aharonian:2006qwBIS}) gamma rays. The X-ray, LE and VHE modulations are in phase, with a peak at inferior conjunction (when the compact object passes in front of the massive star as seen by the observer) and a minimum close to superior conjunction (compact object behind the star). In contrast, the HE gamma-ray modulation is in anti-phase, peaking at superior conjunction. LS 5039's short orbital period of 3.9 days has made it easier to establish these modulations than in the other systems. The modulations appear stable over time, with no reported change in the orbital lightcurves.

Nearly all models invoke synchrotron and inverse Compton emission from pairs with energies up to several TeV to explain the high-energy emission from LS 5039. These processes are efficient in the sense that the energy loss timescale is usually short compared to the flow timescale for the most energetic pairs: a large fraction of the available power may thus end up as high-energy radiation \citep[e.g.][]{Bosch-Ramon:2008hg}. 

The seed photons for inverse Compton emission are provided by the massive star. The pairs upscatter these stellar UV photons to HE and VHE gamma rays. The gamma-ray emission is maximum when pairs backscatter the stellar photons in the direction of the observer i.e. at superior conjunction. However, the gamma-ray photons must propagate through the binary system before reaching the observer. VHE photons are above the threshold for pair production with stellar photons ($\ga 30$\,GeV when interacting with  the $kT_\star\approx 3$\,eV blackbody photons from the star), hence a large part of the VHE flux can be lost to creating $e^{+}e^{-}$ pairs before the radiation escapes. The $\gamma\gamma$ opacity in LS 5039 is minimum close to inferior conjunction and maximum close to superior conjunction, explaining why HE and VHE gamma rays are anti-correlated although both arise from inverse Compton emission of the same seed photons (see \S4 in \citealt{2013A&ARv..21...64D} and references therein). 

There are two caveats to this interpretation of the HE and VHE gamma-ray modulation. First, a straightforward application predicts that none of the VHE flux emitted in the vicinity of the compact object should make it through the system at superior conjunction, whereas observations still detect a faint source. One explanation is that there is a contribution to the emission from the electromagnetic cascade that occurs as newly-created $e^{+}e^{-}$ pairs emit VHE photons which in turn create pairs etc \citep[e.g.][]{Bednarek:2006ka,Bosch-Ramon:2008vd,2010A&A...519A..81C}. Another, non-exclusive, explanation is that the VHE emission arises from a larger or more distant region, diluting the effets of the $\gamma\gamma$ opacity \citep[e.g.][]{2012arXiv1212.3222Z}. The second caveat is that the HE and VHE spectra are very distinct, with the HE spectrum cutting off exponentially at a few GeV \citep{2012ApJ...749...54H}. Clearly, different populations of electrons must be involved in the HE and VHE domains. Their origin is uncertain. Possible sites that have been considered include: the pulsar magnetosphere, the pulsar wind, various locations along the pulsar wind termination shock, the stellar wind termination shock. 

The interpretation of the X-ray and LE gamma-ray modulation also requires additional ingredients. Synchrotron emission dominates in this range. Its luminosity depends on the density of electrons and the magnetic field $B$, both of which can be affected by the changing distance of the termination shock to the pulsar. Indeed, with an eccentricity $e=0.35$ \citep{Casares:2005gg}, the orbital separation $d$ in LS 5039 varies from 0.1 AU at periastron to 0.2 AU at apastron. If the winds are isotropic and have a constant velocity, the distance $R_{s}$ to the termination shock is set by (e.g. \citealt{1990FlDy...25..629L})
\begin{equation}
\frac{R_{s}}{d}=\frac{\eta^{1/2}}{1+\eta^{1/2}},
\end{equation}
where $\eta=\dot{E}/\dot{M}_{w}v_{w}c$, with $\dot{E}$ the pulsar spindown power, $\dot{M}_{w}$ and $v_{w}$ the stellar wind mass loss rate and velocity. The termination shock distance $R_{s}$ doubles from periastron to apastron, decreasing $B\propto 1/R_{s}$ proportionally \citep[e.g.][]{Dubus:2006lc,2009ApJ...702..100T}. The changing separation can also affect adiabatic cooling of the particles as they are advected away from their acceleration site \citep{Takahashi:2008vu}. However, it is difficult to tie such changes to the observed X-ray modulation: the peaks and dips at conjunctions suggest that it is due to a geometrical effect related to the observer line-of-sight rather than due to intrinsic changes in the conditions experienced by the particles. One possibility related to the line-of-sight is Doppler boosting. If $\eta$ is small, the termination shock has the appearance of a bow shock facing away from the massive star. At inferior conjunction the bow shock flows in the general direction of the observer, whereas it is flows in the opposite direction at superior conjunction. If the shocked pulsar wind retains a moderately-relativistic bulk motion, the synchrotron emission will be boosted at inferior conjunction and deboosted at superior conjunction. \citet{2010A&A...516A..18D} have shown that the effect is significant enough to be able to explain the X-ray modulation.

A more accurate assessment of this scenario requires numerical simulations. The geometry of the termination shock, the bulk velocity of the shocked fluid at each location, the adiabatic losses can all be derived from a relativistic hydrodynamical simulation instead of being parametrized as done in previous works. The impact of $\gamma\gamma$ absorption, Doppler boosting and particle cooling on the orbital lightcurve can then be quantified properly, tightening the constrains on the underlying particles and pulsar wind physics. Simulations of gamma-ray binaries have been performed by several groups, focusing on the geometry of the interaction region. Relativistic hydrodynamical simulations indicate that the material re-accelerates to very high Lorentz factors in the tail of the bow shock \citep{2008MNRAS.387...63B} and that the non-zero thermal pressure in the stellar wind leads to smaller opening angles for the bow shock than usually assumed \citep{2013A&A...560A..79L}.  Pulsar wind nebulae are thought to have  a low magnetisation $\sigma$ at the termination shock \citep{Gaensler:2006qi}, in which case the magnetic field has a negligible impact on the shocked flow, and indeed \citet{2012MNRAS.419.3426B}  found little difference, on scales of order of the orbital separation, between their relativistic magneto-hydrodynamical (RMHD) and relativistic hydrodynamical (RHD) simulations of colliding winds in gamma-ray binaries. Other simulation work includes \citet{2012A&A...544A..59B}, who studied mixing due to orbital motion using large scale 2D relativistic hydrodynamical simulations, \citet{2015A&A...574A..77P} who looked at the impact of a clumpy stellar wind on the shock structure, and \citet{2012ApJ...750...70T} who presented two 3D SPH non-relativistic simulations of the interaction of a pulsar wind with a Be disc and wind (covering very large scales with limited spatial resolution).  \citet{2012ApJ...750...70T} computed some lightcurves and spectra from their simulations, but do not take into account particle cooling and relativistic effects. A comprehensive approach linking high-resolution  simulations of the shock region with particle cooling and emission (including relativity and anisotropic effects) has not been attempted yet.

Here, we use a relativistic hydrodynamical simulation as the basis for such a comprehensive model of the emission from particles in the shocked pulsar wind, with the aim of explaining the spectral energy distribution and the flux orbital modulations observed from LS 5039. The simulation (\S2) provides the spatial evolution of the density, velocity and internal energy necessary to compute the extension of the emission, the impact of Doppler boosting, the importance of adiabatic cooling. This radiative post-processing step is described in \S3. The general results are presented in \S4, the more detailed application to LS 5039 in \S5. 

\section{Relativistic hydrodynamics}
We perform a 3D simulation of LS 5039 using the RAMSES hydrodynamical code \citep{2002A&A...385..337T}. The complete description of the extension to special relativity is in \citet{2013A&A...560A..79L}. Here, we recall the major aspects of the method. RAMSES uses a Cartesian grid and allows Adaptive Mesh Refinement (AMR) so as to locally increase the resolution at a reasonable computational cost. It solves the 3D-RHD equations using an upwind second order Godunov method. These equations can be written, for an ideal fluid, as a system of conservation equations in the laboratory frame \citep{Landau}
\begin{eqnarray}\label{eq:RHD}
\frac{\partial{D}}{\partial{t}}+\frac{\partial{(Dv_k)}}{\partial{x_k}} &=&0\\ \label{eq:rel1}
\frac{\partial{M_j}}{\partial{t}}+\frac{\partial{(M_jv_k+p\delta^{j,k})}}{\partial{x_k}} &=&0\\ \label{eq:rel2}
\frac{\partial{E}}{\partial{t}}+\frac{\partial{(E+p)v_k}}{\partial{x_k}} &=&0 \label{eq:rel3},
\end{eqnarray}
where the vector of conservative variables is given by 
\begin{equation}\label{eq:cons_prim}
 \mathbf{U}=
 \begin{pmatrix} 
D \\ 
M_j\\
E 
\end{pmatrix}
=
\begin{pmatrix}
\Gamma \rho \\ 
\Gamma^2 \rho h v_j\\ 
\Gamma^2\rho c^2 h -p
\end{pmatrix}
.
\end{equation}
D is the density, $\mathbf{M}$ the momentum density and E the energy density in the frame of the laboratory. $c$ is the speed of light,  the subscripts $j,k$ stand for the dimensions, $\delta^{jk}$  is the Kronecker symbol. $h$ is the specific enthalpy, $\rho$ is the proper mass density, $v_j$ is the fluid three-velocity, $p$ is the gas pressure. The gravitational force is neglected since we do not treat the acceleration of the winds and the outflow speeds are well above the escape velocity of the system.
The Lorentz factor is given by
\begin{equation}\label{eq:Lorentz}
\Gamma=\frac{1}{\sqrt{1-v^2/c^2}}.
\end{equation}
The set of Eqs. \ref{eq:RHD}-4 is completed by an equation of state that describes the thermodynamics of the fluid. In RHD, the ideal gas approximation is inconsistent with the kinetic theory of relativistic gases \citep{1948PhRv...74..328T}. The exact equation of state for relativistic fluids involves modified Bessel functions, which lead to additional computational costs. Therefore we use the approximation developed by \citet{2005ApJS..160..199M}, which differs from the exact solution by less than 4 $\%$.  In this case, the specific enthalpy is given by 
\begin{equation}
  \label{eq:enthalpy}
  h=\frac{5}{2}\frac{p}{\rho c^2} +\sqrt{\frac{9}{4} \left(\frac{p}{\rho c^2}\right)^2+1}.
\end{equation}
With the specific internal energy defined as $\epsilon=p/\left[(\hat{\gamma}-1)\rho c^2\right]$, this gives a variable equivalent adiabatic index of
\begin{equation}
  \label{eq:ad_index}
  \hat{\gamma}=\frac{h-1}{h-1-\frac{p}{\rho c^2}}.
\end{equation}

In our simulation, we use a HLL solver with a second order reconstruction based on a $minmod$ slope limiter. This solver limits the development of Kelvin-Helmholtz instabilities in the simulation \citep{2013A&A...560A..79L}, a deliberate choice that we justify in \S\ref{p:struct}. The size of our cubic simulation box $l_{\rm box}$ is six times the binary separation $d$.  With $z$ the binary axis, the box extends along the binary axis from $z=-1.5$ to $z=4.5$ in a coordinate system scaled by the orbital separation $d$.  The star is at the origin and the pulsar at $x=0, y=0, z=1$. The $x$-axis extends from $-3$ to $3$ and the $y$-axis extends from $y=-1.5$ to $y=4.5$, so the binary is not located at an edge of the simulation box. Our coarse grid is made of 128 cells and we use four levels of refinement. This gives an equivalent resolution of $2048^3$ cells. The high resolution zone, where the four levels of refinement are effectively used, covers a slab of width $\Delta y=\Delta z=2$ centered on the pulsar and $\Delta z=1$ centered on $x=0$.  Refinement is based on density and Lorentz factor gradients.

We initialize the winds in a spherical region with a continuous outflow. The spherical regions both have a { radius} of $0.15$ (in units of $d$). This is large enough to yield spherical symmetry, but small enough to avoid an impact on the formation of the shock. The winds are updated in the spherical region at every timestep. We assume the winds have reached terminal velocity at launch. A constant low-density medium initially fills the simulation box. This medium is cleared out as the winds propagate out and interact. After a transition phase corresponding to about 10\,$t_{\rm dyn}$ (where we define $t_{\rm dyn}$ as the time for the pulsar wind to reach the back shock located at $x\approx 3$), the simulation converges to a laminar, stationary state where the position of the shocks does not evolve with time anymore {\em i.e.} time drops out of Eq.~2-4. Our radiative calculation is based on a snapshot in the $(y,z)$ plane of the hydrodynamics in this state. The geometry of the interaction depends only on the ratio of wind momentum fluxes $\eta$ (\S1). For LS 5039, $\eta$ is thought to be $\simeq 0.1$, implying a reasonable value for the pulsar spindown power $\dot{E}\approx 4\times 10^{36}\rm\,erg\,s^{-1}$ { assuming} $\dot{M}_{w}\approx 10^{-7}\rm\,M_{\odot}\,yr^{-1}$, $v_{w}\approx 2000\rm\,km\,s^{-1}$ \citep{2011MNRAS.411..193S,2011ApJ...743....7Z}. 

In the simulation, the stellar wind has a mass loss rate of $\dot{M}_w=2\times 10^{-8}\rm\,M_{\odot}\,yr^{-1}$ and a velocity of $2000\rm\,km\,s^{-1}$. It has a Mach number $\mathcal{M}=30$ at the location of the pulsar. Conventional models of pulsar winds invoke ultra-relativistic Lorentz factors up to $\Gamma_p=10^{6}$. Such high values are well beyond the range that can be achieved by standard fluid methods. We set the velocity of the pulsar wind to $v_{p}=0.99$c, or $\Gamma_p=7.08$. This is high enough to capture the relativistic effects in the shocked flow, especially near the apex where the shock is perpendicular and the post-shock flow speed tends to $(\hat{\gamma}-1)c$ when $\Gamma_p\gg 1$. In the wings, the shocked flow gradually accelerates to a velocity close to the initial pre-shock velocity. The spatial scale on which this occurs increases with the initial Lorentz factor  of the wind \citep{2013A&A...560A..79L}. The  pulsar mass loss rate is scaled to have $\eta=0.1$, so that 
\begin{equation}
  \label{eq:pulsar_mass_loss}
  \dot{M_p}=\eta\frac{\dot{M}_w v_w}{\Gamma_p v_p}.
\end{equation}
The flow dynamics and geometry will be correct even if the density must be scaled. For reference, the pulsar wind power corresponding to our choice of $\dot{M}_w=2\times 10^{-8}\rm\,M_{\odot}\,yr^{-1}$ and $\eta=0.1$ in the simulation is $\dot{E}=\Gamma_p \dot{M}_p c^2\approx 7.6\times 10^{35}\rm\,erg\,s^{-1}$.  However, note that we can scale the spindown power up or down without changing the hydrodynamical simulation as long as $\dot{M}_w$ is scaled in proportion to keep $\eta=0.1$. The classical Mach number of the pulsar wind is set to 20, which gives a relativistic Mach number $\mathcal{M}_{\rm rel}=\mathcal{M}\Gamma_p\simeq 140$. Orbital motion turns the shocked structure into a spiral of stepsize of order $S\ga v_s P_{\rm orb}\simeq 4$ AU $\simeq$ 20 $d$ \citep{2012A&A...546A..60L}. Our simulation covers a smaller size ($\simeq 5\,d$) and  we make the assumption that we can neglect orbital motion on this scale\footnote{Neglecting orbital motion makes the problem 2D axisymmetric around the binary axis. RAMSES currently does not allow 2D axisymmetric calculations so our simulation needed to be 3D to allow a quantitative comparison with the observations.}. Even if previous simulations indicate that this is a reasonable approximation when the weaker wind is also the fastest (see {\em e.g.} Fig. 6 in \citealt{2012A&A...546A..60L} or Fig. 1 in \citealt{2012A&A...544A..59B}), a more realistic model should include the orbital motion. We discuss in \S6.2 how this might change our results.

The ratio $\eta$ does not change along the orbit since the winds are isotropic and are at their terminal velocity. The dynamical timescale corresponds to $t_{\rm dyn}\approx 300\,\rm s$ for the typical orbital separation of 0.2\,AU appropriate to LS 5039 (see above). The time taken to form the interaction region, on the scales that we consider, is very short compared to the orbital timescale of 3.9 days for LS 5039. The structure has ample time to reach a stationary state at each orbital phase. Hence, since the stationary structure only depends on $\eta$, which is constant with orbital phase, one simulation is enough for all orbital phases even if the orbital separation changes. The velocities, $\dot{M}$ and wind temperature (Mach number) remain constant. However, for each orbital phase we rescale the distances to the actual orbital separation {\em i.e.} by a factor $d$. The density and pressure in a given pixel are thus rescaled by a factor $1/d^{2}$.

By doing a RHD simulation instead of a RMHD simulation, we implicitely assume that the magnetic field has no dynamical role in the interaction on the spatial scales probed by the simulation. This assumption is supported by the study of \citet{2012MNRAS.419.3426B}, who carried out both relativistic hydrodynamical and relativistic MHD simulations of interacting winds, on spatial scales comparable to those studied here, and found little difference between the two when the pulsar magnetisation $\sigma\leq 0.1$. Conventional models of pulsar wind nebulae assume $\sigma\ll 1$ at the termination shock to ensure the efficient conversion of the wind kinetic energy to particle energy at the shock \citep{Kennel:1984gu,Gaensler:2006qi,2011MNRAS.410..381B}. We come back to this issue in \S6.2.

\section{High-energy radiation\label{p:he}}

\begin{figure}
\centerline{\includegraphics[width=\linewidth]{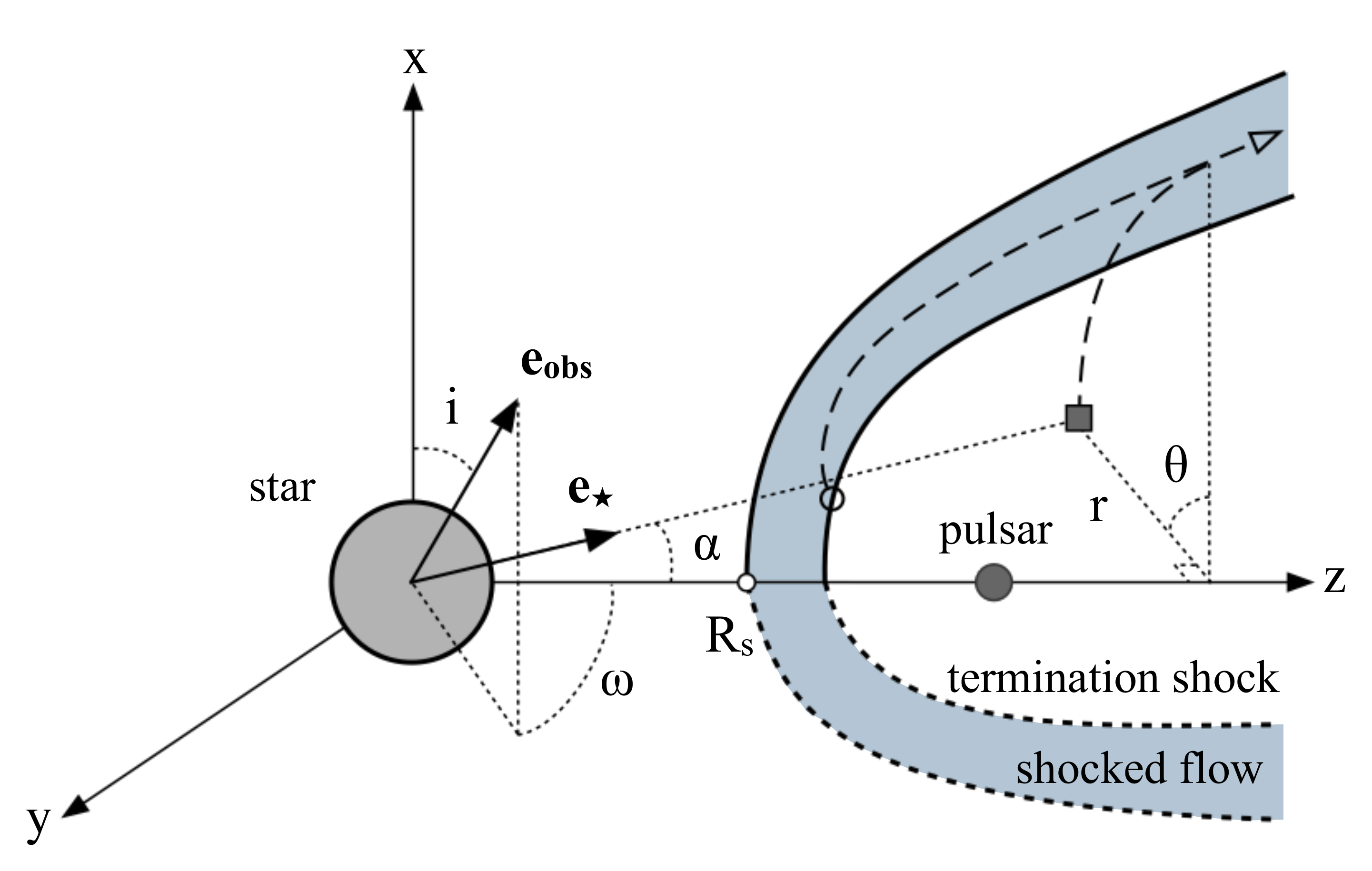}}
\caption{Model geometry showing our choices of axis and angles. The interaction has cylindrical symmetry around the $z$ axis. The two thick solid lines in the $(x,z)$ plane represent the pulsar wind termination shock and the contact discontinuity with the shocked stellar wind. Dashed lines represent the same in the $(y,z)$ plane. Particles injected at the termination shock follow streamlines in the region in between these two surfaces.}
\label{fig:geometry}
\end{figure}

The interaction of stellar winds classically leads to a double shock structure separated by a contact discontinuity (Fig.~\ref{fig:geometry}). Non-thermal emission may be expected from the shock associated with both winds \citep{2011MNRAS.418L..49B}. There is plenty of evidence for efficient high-energy non-thermal emission from pulsar wind termination shocks, but rather less from stellar wind termination shocks with only Eta Carinae detected so far in gamma rays \citep{2013A&A...555A.102W}. Here, we have assumed that the non-thermal emission results exclusively from particles accelerated at the pulsar wind termination shock. We have not taken into account thermal bremsstrahlung from the shocked stellar wind, since this would be best treated by dedicated simulations {\em \`a la} \citet{Stevens:1992on}, and because the X-ray emission from LS 5039 appears entirely non-thermal with, as yet, no evidence for thermal emission \citep{2011ApJ...743....7Z}.

Particles are randomized and/or accelerated to very high energies at the termination shock of the pulsar. We wish to follow the evolution of these particles once they are injected in the shocked pulsar wind flow. Once their energy distribution at each location is known, we can compute the radiation seen at a given line-of-sight to the binary.

We adopted some basic assumptions to make the problem tractable. First, the evolution of the particles is decoupled from the hydrodynamical simulation and treated in the test particle limit as a post-processing step {\em i.e.} radiative cooling does not impact the dynamics of the flow.  We come back to this in \S\ref{sec:cooling} and \S5.1. Second, the high-energy particles are assumed to follow the flow. Spatial diffusion is expected to remain negligible if the Larmor radius remains small compared to the flow spatial scales, which is the case here. Third, plasma processes are assumed to keep the particle momentum distribution isotropic. Last, the flow is assumed to be stationary, simplifying the problem to that of following the evolution of particles along streamlines. 

We start by explaining how we calculate the total emission based on the particle evolution along streamlines (\S\ref{sec:total}). We then describe how we choose the initial distribution of the injected particles (\S3.2), how we compute the particle evolution and streamline emission (\S\ref{sec:stream}), how we estimate the magnetic field (\S\ref{sec:mag}), and how we deal with the changing aspect due to orbital motion (\S\ref{sec:geometry}).

\subsection{Total flux\label{sec:total}}
The total flux from the emission region, measured in the laboratory frame at a frequency $\nu$, is given by 
\begin{equation}
F(\nu)=\frac{1}{D^2} \int_V \mathcal{D}_{\rm obs}^2  j \left({\nu}/{\mathcal{D_{\rm obs}}}\right) \exp\left(-\tau_\nu\right)dV,
\label{eq:flux}
\end{equation}
where the integral is over the volume $V$ of the region (measured in the laboratory frame), $D$ is the distance to the source, $j$ is the local particle emissivity in the co-moving frame ($j$ is the sum of the synchrotron $j_{\rm sync}$ and inverse Compton  $j_{\rm ic}$ emissivities), and $\tau_\nu$ is the opacity at frequency $\nu$ due to pair production as the photons emitted in the volume $dV$ travel along the line-of-sight to the observer (see \S\ref{sec:stream}).  $\mathcal{D}_{\rm obs}$ is the Doppler boost to the observer
\begin{equation}
\mathcal{D}_{\rm obs}=\left[\Gamma \left(1-\mathbf{v}.\mathbf{e_{\rm obs}}\right)\right]^{-1},
\label{eq:dobs}
\end{equation}
with $\mathbf{v}$ the flow velocity vector and $\mathbf{e_{\rm obs}}$ the unit vector in the direction of the observer (\S\ref{sec:geometry}).

The simulated flow is laminar. To compute this integral, we extract from a snapshot $\cal N$ streamlines in the emission region. Each streamline starts at the position of the shock and ends where it leaves the computational domain. These streamlines subdivide the emission region into streamtubes\footnote{Rather than streamtubes, these are actually hollow cones since the model has axial symmetry around the binary axis, see Fig.~\ref{fig:geometry}.}. The integral over the volume can be written as an integral over time of the particles flowing along each streamtube:
\begin{equation}
F(\nu)=\frac{1}{D^2}\sum_{i=1}^{\cal N} \int_0^{t_i}  \mathcal{D}_{\rm obs}^2 j\left(\nu/\mathcal{D}_{\rm obs}\right)  \exp\left(-\tau_\nu\right) \mathbf{v}.\mathbf{S} dt,
\label{eq:flux1a}
\end{equation}
where $\mathbf{S}$ is a cross section of the streamtube $i$ and $t=t_i$ is the time taken by particles flowing along streamline $i$ to leave the computational domain, with $t=0$ at their injection at the shock. Since the flow is stationary, the particle flux along each streamtube 
\begin{equation}
\dot{N}_{i}\equiv \left.\Gamma n \mathbf{v}.\mathbf{S}\right|_{i}
\label{eq:pflux}
\end{equation}
 is constant ($n=\rho/m_{e}$ in the pulsar wind composed of $e^{+}e^{-}$ pairs). The total flux from the shocked region becomes
\begin{equation}
F(\nu)=\frac{1}{D^2}\sum_{i=1}^{\cal N}  f_i(\nu) \dot{N}_{i},
\label{eq:flux2}
\end{equation}
where we have defined the fluence per particle $f_i$ of the streamline $i$ as 
\begin{equation}
f_i(\nu)\equiv \int_0^{t_i} \frac{ \mathcal{D}^2_{\rm obs}  j \left({\nu}/{\mathcal{D_{\rm obs}}}\right)\exp\left(-\tau_\nu\right)} {\Gamma n} dt.
\label{eq:intstream}
\end{equation}
This is an integral over the emission per particle as they flow along their streamline. The integral can also be recast as an integral over position along the streamline since the curvilinear distance $l$ is related to time by $dl=v dt$. We obtain $\dot{N}_i$ numerically by measuring the particle flux across the shock surface for each streamtube in the RHD simulation (Eq.~\ref{eq:pflux}).  

We make the implicit assumption in Eq.~\ref{eq:flux2} that the emissivity $j$ and the opacity $\tau_\nu$ do not  vary substantially across the streamtube for a given position. In practice, taking a sufficient number of streamlines ensures this is achieved. We used ${\cal N}=920$ streamlines, evenly spread along the shock surface, and we checked that this is more than enough for numerical convergence of  Eq.~\ref{eq:flux2}.

Alternatively, one could divide up the particle distribution into energy bins, recast the evolution equation of the particles in each energy bin in conservative form and add them to the set solved by RAMSES \citep[e.g.][]{2014ApJ...782...96R}. The advantage is that this can deal with non-stationary flows and mixing. However, a major drawback is that in our case the particle cooling time at very high energies can be very short: the particles cool on small spatial scales, requiring a very high spatial resolution (see \S\ref{sec:cooling} below). Another difficulty is that radiative cooling does not scale with the orbital separation $d$ as adiabatic cooling ($d^{-2}$ and $d^{-1}$ respectively, see Eq.~\ref{eq:evol} below). Each orbital phase would thus require a full simulation to account for this difference in cooling spatial length.  Each new choice of parameters for radiative cooling would also require a new simulation run. Treating the particle evolution in a post-processing step makes the problem much more tractable, allowing for a wider exploration of the parameter space involved in the flow emission.

We explain our choices for particle injection at the shock first before turning to the computation of the streamline fluence (Eq.~\ref{eq:intstream}) in \S\ref{sec:stream}.

\subsection{Particle injection at the shock}
We distribute the particles as a power-law function of their Lorentz factor $\gamma$:
\begin{equation}
\label{eq:pl}
\left. \frac{dn}{d\gamma}\right|_{\rm t=0} = K \gamma^{-s}  ~~{\rm with}~~\gamma_{\rm min}\leq\gamma\leq \gamma_{\rm max}.
\end{equation}
The particle density at the shock sets the normalisation $K$ of the distribution. The two other parameters are $\gamma_{\rm min}$ and $\gamma_{\rm max}$.

The maximum Lorentz factor $\gamma_{\rm max}$ is set by the balance between acceleration and radiative losses, since the gyroradius will be much smaller than the characteristic size of the acceleration region in our case \citep{Dubus:2006lc}. We assume that the acceleration timescale in the comoving frame $\tau_{\rm acc}$  is some multiple of the B\"ohm limit: 
\begin{equation}
\tau_{\rm acc}=2\pi \xi R_L/c,
\label{eq:bohm}
\end{equation}
with $R_{L}$ the gyroradius. We expect  $\xi\geq 1$ for diffusive shock acceleration, with $\xi\leq 10$ corresponding to ``extreme acceleration'' \citep{Khangulyan:2007me} ;  $\xi < 1$ may be possible for acceleration at magnetic reconnection sites, as proposed to explain the gamma-ray flares from the Crab nebula \citep{2012ApJ...746..148C}. Synchrotron losses dominate over inverse Compton losses at very high energies \citep{Dubus:2006lc}. The synchrotron loss timescale is 
\begin{equation}
\tau_{\rm sync}\equiv\gamma\left(\frac{d\gamma}{dt^\prime}\right)^{-1}=\frac{\gamma m_e c^2}{\frac{4}{3}\sigma_T c (\beta \gamma)^2 u_b}\approx 77\, \left(\frac{1\,{\rm G}}{b}\right)^{2} \left(\frac{10^7}{\gamma}\right)~{\rm s},
\label{eq:sync}
\end{equation}
where $b$ and $t^{\prime}$ are the magnetic field and time in the comoving frame. Setting $\tau_{\rm acc}\leq\tau_{\rm sync}$ gives
\begin{equation}
\gamma_{\rm max}=\left(\frac{3 e }{\xi  \sigma_T b}\right)^{1/2}\approx 5\times 10^7\  \xi^{-1/2} \left(\frac{1\,{\rm G}}{b}\right)^{1/2}.
\label{eq:gmax}
\end{equation}

With $\gamma_{\rm max}$ and $K$ known, we derive $\gamma_{\rm min}$ by writing that the energy in the particle distribution is a fraction $\zeta_{p}$ of the downstream specific energy $\epsilon$:
\begin{equation}
\zeta_p\epsilon\equiv \epsilon_{\rm nt} = \left. \left(\int_{\gamma_{\rm min}}^{\gamma_{\rm max}} \gamma m_{e}c^{2} \frac{dn}{d\gamma}d\gamma\right) ~ \middle/ ~\left(\int_{\gamma_{\rm min}}^{\gamma_{\rm max}} \frac{dn}{d\gamma}d\gamma\right)\right. .
\label{eq:gmin}
\end{equation}
The numerator is the total energy in the distribution and the denominator is the particle density. This equation implicitly sets $\gamma_{\rm min}$, which can be found numerically with a Newton-Raphson scheme. If $\gamma_{\rm max}\gg \gamma_{\rm min}$ and $s> 2$ then the integrals can be simplified and solved for $\gamma_{\rm min}$, such that
\begin{equation}
\gamma_{\rm min}\approx  \zeta_{p}\epsilon \left(\frac{s-2}{s-1}\right)=\frac{\zeta_{p}}{ (\hat{\gamma}-1)} \left(\frac{s-2}{s-1}\right) \frac{p}{ n m_e c^{2}}.
\end{equation}
$\zeta_p$ enables us to scale the emission to realistic values since the (lab-frame) specific enthalpy of the cold pulsar wind at the shock is $\Gamma_p\approx 7$ in the simulation, which is too low to reproduce the very high values of $\epsilon$ (equivalently $\Gamma_p$) required to fit the observations. Raising $\zeta_p$ implies that the effective mass loss rate from the pulsar wind is decreased by $\zeta_p^{-1}$ because $\dot{E}/c=\zeta_p \Gamma_p \dot{M}_p c$ is fixed by $\eta$ for given $\dot{M}_w$, $v_w$, $\Gamma_p$ (in other words, the same total available energy is distributed amongst fewer particles).

We have also experimented with a relativistic Maxwellian distribution because {\em Fermi}-LAT observations of gamma-ray binaries in HE gamma rays require an additional population of particles with a narrow range in energy (\S1). Moreover, particle-in-cell simulations of relativistically-shocked pair plasmas typically show a prominent Maxwellian distribution of shock-heated particles together with the power-law distribution of accelerated particles \citep{2009ApJ...698.1523S,2011ApJ...741...39S}. The relativistic Maxwellian distribution is 
\begin{equation}
\left. \frac{dn}{d\gamma}\right|_{\rm t=0}\equiv K \gamma^{2}\exp(-\gamma/\gamma_{t}).
\end{equation}
Again, $K$ is derived by imposing that the integral of the distribution scales with the particle density at the shock. The mean Lorentz factor of the distribution  $\gamma_{t}$ is derived from Eq.~\ref{eq:gmin}, which gives $\gamma_{t}\approx e_{\rm nt}/3$ when $\gamma_{\rm min}\ll\gamma_{t}\ll \gamma_{\rm max}$. 

We assumed that $\zeta_p$ and $\xi$ do not vary along the shock for lack of strong justifications for more sophisticated assumptions.

\subsection{Streamline emission\label{sec:stream}}
For a stationary flow, the evolution of a particle injected at a given location and followed along the associated streamline in the comoving frame, is uniquely set by the evolution equation \citep[e.g.][for applications to pulsar wind nebulae or AGN jets]{Del-Zanna:2006rr,2009ApJ...696.1142M,2014MNRAS.438..278P} 
\begin{equation}
\frac{1}{\gamma}\frac{d\gamma}{dt^\prime}=\frac{d\ln \epsilon}{dt^\prime}-\frac{1}{\tau_{\rm sync}}-\frac{1}{\tau_{\rm ic}},
\label{eq:evol}
\end{equation}
where the terms on the right hand side represent adiabatic, synchrotron and inverse Compton losses -- the only relevant cooling processes here. The elapsed time in the laboratory frame $dt$ is related to the proper time in the comoving frame by $dt=\Gamma dt^\prime$. The bulk Lorentz factor $\Gamma$ and the elapsed time $dt$ are derived from the position and velocity along the streamline as calculated with RAMSES.  The adiabatic loss term is also directly derived from the simulation.

The evolution of the magnetic field along the streamline must be known to compute the synchrotron losses  $\tau_{\rm sync}$ (Eq.~\ref{eq:sync}) and the associated emissivity $j_{\rm sync}$. Our choices for the magnetic field are  explained below in \S3.4. The synchrotron emissivity $j_{\rm sync}$ is computed using the usual formula involving Bessel functions \citep{Rybicki:1979za}.

For the inverse Compton losses $\tau_{\rm ic}$, we take the massive star as the only source of seed photons and compute the electron energy losses $\tau_{\rm ic}$ using the \citet{Jones:1968pe} scattering kernel.  The star is modeled as a blackbody of temperature $T_\star=39\,000 K$ and radius $R_\star=9.3 R_{\odot}$. The density $n_\star$ of photons with an energy $\epsilon_\star$ (in units of $m_e c^2$) seen in the comoving frame by pairs at a distance $d_{\star}$ from the star is, in photons per cm$^3$ per unit energy, 
\begin{equation}
n_\star=2\pi\left(\frac{m c}{h}\right)^3 \left(\frac{R_\star}{d_\star}\right)^2 \frac{\epsilon_\star^2/\mathcal{D}_\star^2 }{\exp\left(\frac{\epsilon_\star m_e c^2}{\mathcal{D}_\star kT_\star}\right)-1}.
\end{equation}
$\mathcal{D}_\star$ is the Doppler boost required to transform the stellar radiation field into the comoving frame,
\begin{equation}
\mathcal{D}_\star=\left[\Gamma \left(1-{\mathbf{v}.\mathbf{e_{\star}}}\right)\right]^{-1},
\label{eq:dstar}
\end{equation}
with $\mathbf{e_{\star}}$ the unit vector giving the direction from the star to the flow element containing the pairs. 

The evolution of the particle distribution can be calculated semi-analytically when inverse Compton losses are in the Thomson regime \citep{Begelman:1992gp}. A numerical solution is required in our case since stellar photons are upscattered to gamma-ray energies in the Klein-Nishina regime \citep{Dubus:2007oq}. Following \citet{Bosnjak:2008mb}, the particle distribution of each streamline is discretized in ``Lagrangian" bins 
\begin{equation}
n_k=\int_{\gamma_k}^{\gamma_{k+1}} \frac{dn}{d\gamma} d\gamma~~{\rm with}~~\sum_k n_k= n.
\end{equation}
The relative number of particles $n_k/n$ in each energy bin is conserved but the bin boundaries $[\gamma_k,\gamma_{k+1}]$ vary along the streamline according to the energy loss equation (Eq.~\ref{eq:evol}). We use 400 energy bins, initially logarithmically distributed between $\gamma_{\rm min}$ and $\gamma_{\rm max}$. For each $\gamma_{k}$, Eq.~\ref{eq:evol} is integrated in time steps representing a small fraction (5\%) of the minimum energy loss timescale. To ease the computational burden, we stop following the energy losses once $\gamma$ becomes lower than $10^{3}$: we verified that these particles do not contribute emission in the frequency range we are interested in.

We compute the streamline fluence $f_i(\nu)$ (Eq.~\ref{eq:intstream}) once the  evolution of the particle distribution along the streamline is known. The fluence depends on the location in the flow but also on the line-of-sight to the observer because of the relativistic boost associated with the bulk motion of the flow ${\cal D}_{\rm obs}$ (Eq.~\ref{eq:dobs}). Unlike the particle evolution calculation, which is symmetric around the binary axis and  requires only a 2D integration (see \S\ref{sec:geometry}), the emission calculation requires a full 3D integration since $\mathbf{e}_{\rm obs}$ also varies with the azimuth $\theta$ around the binary axis (Fig.~\ref{fig:geometry}).

Synchrotron radiation is isotropic in the comoving frame for a random orientation of the magnetic field and an isotropic distribution of particles, both reasonable assumptions at this stage. However, the inverse Compton emission is clearly not isotropic in the comoving frame, even for an isotropic distribution, because the seed photons from the star are anisotropic \footnote{By using the Jones kernel we have assumed that particles (continuously) lose energy isotropically on average. This is consistent with anisotropic emission if plasma processes maintain the particle distribution isotropic.}.  We follow \citet{2010A&A...516A..18D} to take this effect into account when computing the upscattered emissivity $j_{\rm ic}$ towards the observer line-of-sight. Finally, we also calculate the $\gamma\gamma$ absorption of the VHE flux due to pair production with the stellar photons as in \citet{Dubus:2006lr}. The $\gamma\gamma$ opacity $\tau_{\gamma\gamma}$ depends on the path of the VHE photon from its emission location to the observer. We neglected the finite size of the star to ease the computational requirements. This approximation appears reasonable here since the extension of the VHE emission region would probably reduce and smoothe out any effect of the finite size on both the inverse Compton emission and the $\gamma\gamma$ opacity. Knowing $\tau_\gamma$, $j_{\rm ic}$, and $j_{\rm sync}$  along each streamline enables us to evaluate Eq.~\ref{eq:intstream}.

\subsection{Magnetic field\label{sec:mag}}
Following the standard description of pulsar winds, the laboratory frame magnetic field $B_p$ at a distance $d_{p}$ from the pulsar in the unshocked wind is given by
\begin{equation}
\frac{B_p^{2}}{4\pi}\left(\frac{1+\sigma}{\sigma}\right)=\frac{\dot{E}}{4\pi d_{p}^2 c},
\end{equation}
where $\sigma$ is the wind magnetisation
\begin{equation}
\sigma\equiv\frac{B_p^{2}}{4\pi\Gamma^{2}_p n_p m_e v_p^2},
\end{equation}
and the subscript $p$ identifies a value in the unshocked wind. Pulsar winds are thought to have a low $\sigma$ so we neglect the influence of the magnetic field on the flow dynamics (\S2).

The magnetic field is purely toroidal at large distances from the magnetosphere of the pulsar, so the shock is perpendicular and the field is amplified at the shock by the compression ratio $\chi=B/B_p=N/N_p$, where $N=\Gamma n$ is the laboratory frame number density. The compression ratio depends on the adiabatic index $\hat{\gamma}$, which can change along the shock (see Eq.~\ref{eq:enthalpy}). Even if $\hat{\gamma}$ is constant, the compression ratio changes in a relativistic shock when the transverse speed is non-negligible. For a ultra-high wind Lorentz factor, the shock behaves like a normal shock with  $\chi\approx 1/(\hat{\gamma}-1)$ except very far out in the wings, where the speed normal to the shock becomes non-relativistic and $\chi=(\hat{\gamma}+1)/(\hat{\gamma}-1)$. We have taken $\chi=1/(\hat{\gamma}-1)$ everywhere along the shock, so that
\begin{equation}
B=\frac{1}{\hat{\gamma}-1}\left[\frac{\dot{E}}{d_{p}^2 c}\left(\frac{\sigma}{1+\sigma}\right)\right]^{1/2}.
\label{eq:b1}
\end{equation}
In practice we take $B=\zeta_{b}/d_{p}$, where $\zeta_{b}$ is a constant, since $\dot{E}$ and $\sigma$ are not fixed but will be determined by the comparison to data.

Alternatively, the magnetic field could be amplified by plasma instabilities to a value that represents a fraction of the available internal energy
\begin{equation}
\frac{b^{2}}{8\pi}  = \zeta_b \frac{p}{\hat{\gamma}-1},
\label{eq:b2}
\end{equation}
where $b=B/\Gamma$ is the magnetic field in the comoving frame, again assumed to be toroidal, and  (again)  $\zeta_{b}$ is a constant. From the conservation of $\Gamma h$ across the shock, the pressure is related to the upstream conditions by 
$$p=\frac{\hat{\gamma}-1}{\hat{\gamma}}\left(\frac{\Gamma_p}{\Gamma}\right)^2\frac{v_p}{v}\left(1-\frac{\Gamma}{\Gamma_p}\right) n_p m_e c^2 $$
when the upstream pressure is negligible\footnote{When $\Gamma_p\gg1 $ then $\Gamma\approx (\hat{\gamma}-2\hat{\gamma}^2)^{-1/2}$ and  $p\approx \left(2-\hat{\gamma}\right)\Gamma_p N_p m_e c^2 $, showing that the kinetic energy of the wind is tapped.}. This equation shows that $p\propto (\Gamma d_p)^{-2}$, hence, that $B\propto 1/d_p$. Thus, there is little difference in practice between the magnetic field $B$ as given by Eq.~\ref{eq:b1} and Eq.~\ref{eq:b2}, the two being exactly proportional when the pulsar wind Lorentz factor is high ($\Gamma_p\gg 1$).

Since we assume no influence of the magnetic field on the flow, the evolution of $B$ beyond the shock is set solely by the induction equation. For a stationary flow, with cylindrical symmetry around the binary axis, the induction equation for the purely toroidal $B$ field becomes
\begin{equation}
\pd{\mathbf{B}}{t}-\nabla \times ({\mathbf v}\times \mathbf{B}) = \pd{v_r B}{r}+\pd{v_zB}{z}=0,
\end{equation}
with $z$ the coordinate along the symmetry axis and $r$ the radial cylindrical coordinate (Fig.~\ref{fig:geometry}). The induction equation is identical to the continuity equation if $B$ is  replaced by $\Gamma nr$ \citep{1999A&A...349..323M,2012MNRAS.419.3426B}. Hence, the evolution of $B$ along the flow streamline simply follows $B\propto \Gamma nr$, with the proportionality constant set by the initial conditions at the shock.

\subsection{Geometrical factors and orbital motion\label{sec:geometry}}
Taking the origin of the coordinate system at the center of the massive star, $z$ along the binary axis and $r$ the radial coordinate perpendicular to the binary axis then the flow element is at $(r\cos\theta,r\sin\theta,z)$ where $\theta$ is the azimuthal angle around the binary axis of symmetry  (see Fig.~\ref{fig:geometry}). The unit vector $\mathbf{e}_{\star}$ is
\begin{equation}
\mathbf{e}_{\star}=(\sin\alpha \cos\theta, \sin\alpha \sin\theta, \cos\alpha),
\end{equation}
with $\tan \alpha= r/z$. Taking advantage of the symmetry around the binary axis, the speed in the laboratory frame is $\mathbf{v}=(v_r\cos\theta, v_r\sin\theta, v_{z})$. It is straightforward to see that the boost $\mathcal{D}_\star$ (Eq.~\ref{eq:dstar}) that applies to the stellar emission seen in the comoving frame does not depend on $\theta$: the evolution can be computed using a set of streamlines taken in a plane including the binary axis. The unit vector to the observer is 
\begin{equation}
\mathbf{e}_{\rm obs}= (\cos i, \sin\omega\sin i, \cos\omega\sin i),
\end{equation}
with $i$ is the inclination of the system and $\omega$ the true anomaly of the orbit.  

For each orbital phase, the simulation is scaled with the orbital separation $d$:
\begin{equation}
d=\frac{a(1-e^2)}{1-e\sin(\omega-\omega_{p})}.
\end{equation}
The orbital parameters are those of LS~5039 \citep{2004ApJ...600..927M,Casares:2005gg,2011heep.conf..559C} i.e. the semi-major axis $a=(G M P^2_{\rm orb}/4\pi^2)^{1/3}$ with $P_{\rm orb}=3.9$\,days, $M=1.4\,M_{\odot}+23\,M_{\odot}$ the total mass, $e=0.35$ the eccentricity, $\omega$ the true anomaly, and $\omega_{p}=212\degr$ the angle at which the binary is at periastron.  We divide the orbit into 30 phases, each of which requires a (2D) calculation of the particle evolution and a (3D) calculation of the observed emission. We use $20$ cells for $\theta$, 400 for the electron Lorentz factor $\gamma$, 20 for the stellar photon energy $\epsilon_\star$. The calculations were parallelized using OpenMP.

\section{Results}

\subsection{The structure of the shocked flow\label{p:struct}}
\begin{figure}
\centerline{\includegraphics{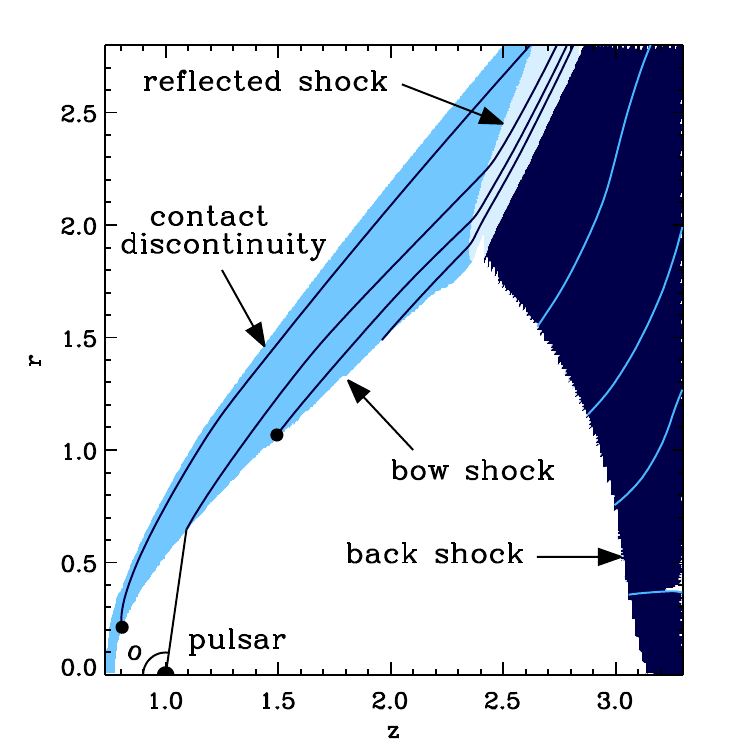}}
\caption{The RHD flow in the shocked pulsar wind. The star is at the origin $(0,0)$ and the pulsar is at $(1,0)$ in units of the orbital separation $d$. A few selected streamlines have been plotted. Three different regions have been colored. They correspond to the {\em bow} shock (blue), the {\em reflected} shock (light blue), and the {\em back} shock (dark blue) regions.}
\label{fig:structure}	
\end{figure}

The numerical simulation shows the expected double shock structure. Numerical diffusivity, induced by our choice of Riemann solver, stabilizes the structure despite the presence of a strong velocity shear at the interface between the shocked pulsar and stellar winds. The diffusivity leads to  gradual mixing between the winds {\em i.e.} numerical spreading of the contact discontinuity, quenching the development of the Kelvin-Helmholtz (KH) instability \citep{2011MNRAS.418.2618L}.  Strong KH mixing could impact the emission of the region, for instance by reducing the Lorentz factor of the flow, and by generating strong turbulence. The fluctuation timescale of the interface would be short since the flow is relativistic. However, the strong velocity shear is accompanied by a  strong density contrast between the dense stellar wind and the tenuous pulsar wind. The ratio of the KH growth timescale to the advection timescale is $\propto (\rho_2/\rho_1)^{1/2} \Delta v $ for two fluids of density $\rho_1$ and $\rho_2\ll \rho_1$, sheared by a velocity difference $\Delta v$ (see appendix in \citealt{2012A&A...546A..60L} and \citealt{2004PhRvE..70c6304B} for the growth rate in the relativistic regime). Hence, the KH growth is dampened for high density contrasts, such as that expected between the tenuous highly relativistic pulsar wind and the dense stellar wind, making it debatable whether KH-induced mixing is dynamically important in gamma-ray binaries on  the scales that we consider here. \citet{2012A&A...544A..59B,2014arXiv1411.7892B} find mixing occurs mostly on larger scales  in their simulations and attribute it rather to instabilities triggered by orbital motion. Large, dense clumps in the stellar wind could also affect the shock structure and variability \citep{2015A&A...574A..77P}, although it is unclear whether there is enough time for the clumps to grow before reaching the termination shock in LS 5039 (located within 1-2 stellar radii of the star). Our simulation implicitely assumes that any mixing is limited and roughly captured --- in a time-averaged sense --- by the numerical diffusivity.

The basic structure of the shocked pulsar wind is illustrated in Fig.~\ref{fig:structure}, where we show only part of the full simulation domain. The stellar wind (shocked and unshocked) and the unshocked pulsar wind have been edited out of this map as well as subsequent ones since our focus is entirely on the shocked pulsar wind. The head of the pulsar wind is shocked in a bow-shaped region with asymptotic angles $\approx 40\degr$ (termination shock) to $50\degr$ (contact discontinuity), measured from the z axis. This is larger than the $30\degr$ to $45\degr$ found  by \citet{2008MNRAS.387...63B} for $\eta=0.1$, but our simulation domain is smaller and our angles may not yet have reached their true asymptotic values. 

Besides the bow-shaped shock, our simulation shows that the pulsar is also terminated at the back instead of propagating freely. The structure is the one expected for Mach reflection on the binary axis as described in \citet{2008MNRAS.387...63B} in the context of gamma-ray binaries and as observed in {\em e.g.} simulations of pulsar bow-shock nebulae \citep{2004ApJ...616..383G}.  Material flowing in the bow shock region abruptly changes direction when it crosses into the light blue region (black streamlines in Fig.~\ref{fig:structure}). The change is due to a reflection shock that appears in order to accommodate the back shock region. This {\em reflected} shock region is separated by a contact discontinuity from the back shock region (boundary between medium and dark blue regions in Fig.~\ref{fig:structure}).  

The back shock structure in our simulation is very similar to the back shock structure in the 2D and 3D relativistic simulations of \citet{2012A&A...544A..59B,2014arXiv1411.7892B}, who use a different code (PLUTO) but similar values for $\eta$ and $\Gamma_p$ ($\eta=0.1$ and $\Gamma_p=2$ for their 3D simulation). All their simulations include orbital motion and they interpret the presence of this structure as an effect of orbital motion. Since this cannot be the case in our simulation, we suspect that the confinement depends on a subtle combination of 3D + relativistic + pressure (Mach number) effects  \citep{2011MNRAS.418.2618L,2012A&A...546A..60L,2013A&A...560A..79L}. We defer a resolution of this possible issue  to future studies. In the present case, as we shall see, the back shock and reflected shock only play a minor role in shaping the high-energy emission.

\begin{figure}
\centerline{\includegraphics{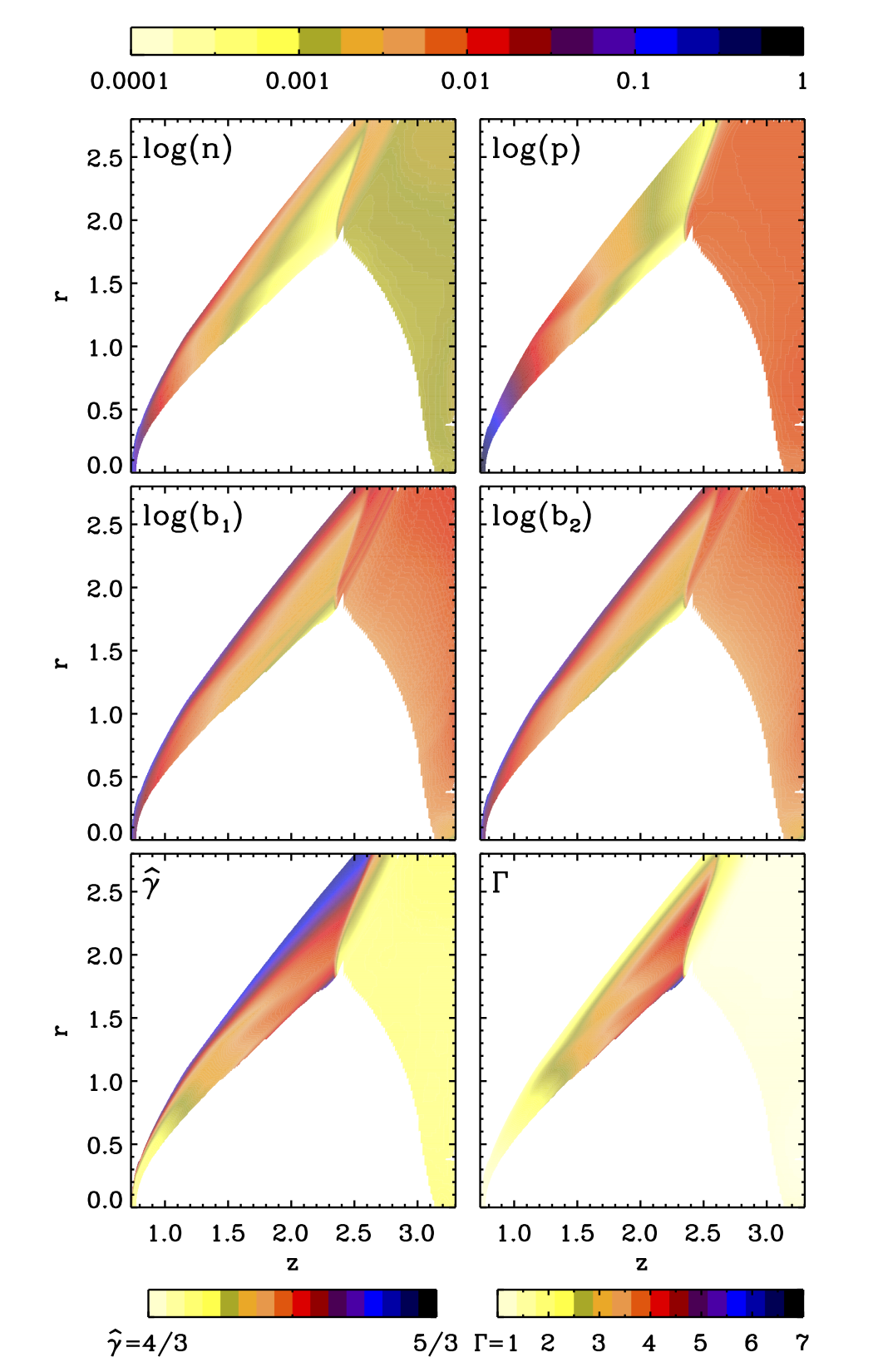}}
\caption{Maps of various quantities in the shocked pulsar wind. The particle density $n$, pressure $p$, and magnetic field are displayed on a logarithmic scale ranging down to $10^{-4}$ of the maximum value (top colorbar).  The magnetic field $b_1$ (resp. $b_2$) is calculated using Eq.~\ref{eq:b1} (resp. Eq.~\ref{eq:b2}) at the shock. The adiabatic index $\hat{\gamma}$ and Lorentz factor $\Gamma$ are displayed on a linear scale (bottom colorbars). The map spatial scale is in units of the orbital separation $d$ with the pulsar at $(1,0)$ and the star at $(0,0)$.}
\label{fig:map_flow}
\end{figure}

Figure~\ref{fig:map_flow} shows maps of the various flow quantities in the shocked pulsar wind. The jumps in density single out the reflected shock region. The jump in pressure identifies the interface with the bow shock flow as a shock while the matching pressures identifies the interface with the back flow as a contact discontinuity. The bow shock flow is re-energized by adiabatic compression when it crosses the reflected shock. The magnetic field distribution is identical regardless of the assumption adopted for $B$ at the pulsar termination shock (\S\ref{sec:mag}). The highest magnetic field intensities are found at the contact discontinuity with the stellar wind, where streamlines from the bow shock head pile up. The magnetic field increases with the density in the reflected shock region ($b\propto n r$). The last two panels show the adiabatic index $\hat{\gamma}$ and the Lorentz factor. The adiabatic index is that of a relativistic gas ($\hat{\gamma}=4/3$) when the shock is perpendicular and decreases towards its non-relativistic value ($\hat{\gamma}=5/3$) as material flows in the bow shock region due to adiabatic expansion. The bow shock flow accelerates  back up to a fraction of the initial Lorentz factor of the free pulsar wind before being slowed down again by the reflected shock. The properties in the back flow region vary little on the scales examined here: the flow speed remains close to $v=c/3$ with $\hat{\gamma}\approx 4/3$, and a slowly varying pressure and density.

\subsection{Particle cooling\label{sec:cooling}}

\begin{figure}
\centerline{\includegraphics{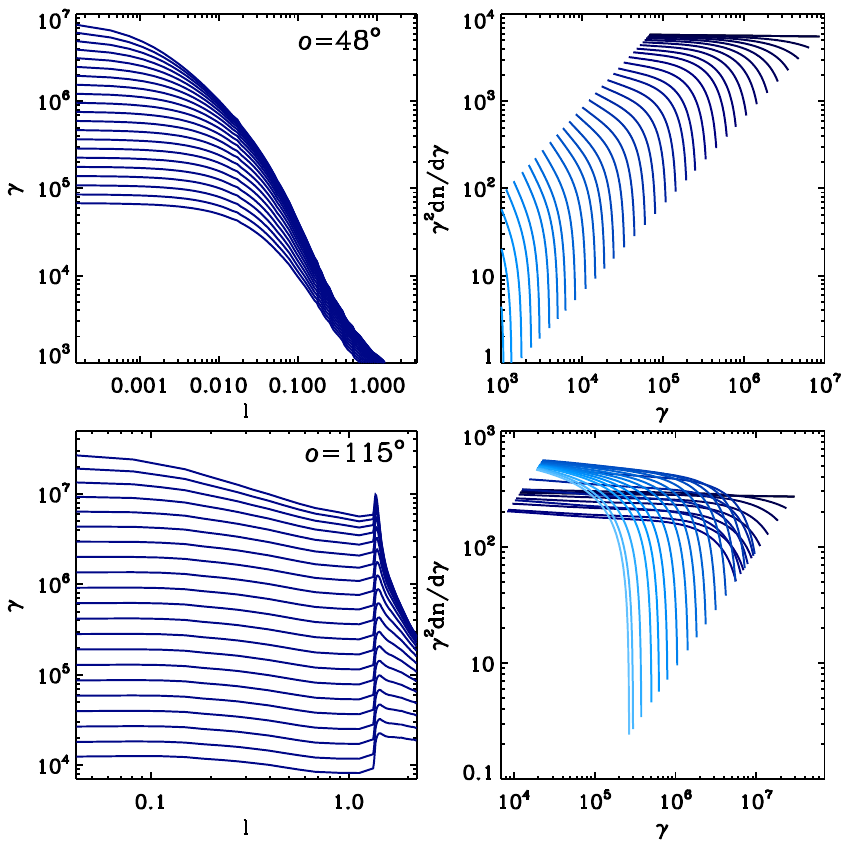}}
\caption{Left: evolution of particle energy as a function of the distance $l$ along the streamline (in units of the orbital separation). Right: evolution of the particle energy distribution along the streamline (lighter colors for later times i.e. increasing distance $l$). The top panels correspond to the streamline starting at $o=48\degr$, the bottom panels to the streamline starting at $o=115\degr$ (both are identified by an initial dot in Fig.~\ref{fig:structure}). The distribution in the bottom right panel ``jumps" when the particles cross the reflected shock and are re-energised. The evolution is calculated at periastron for our reference simulation.}
\label{fig:part_evolution}
\end{figure}
\begin{figure}
\centerline{\includegraphics{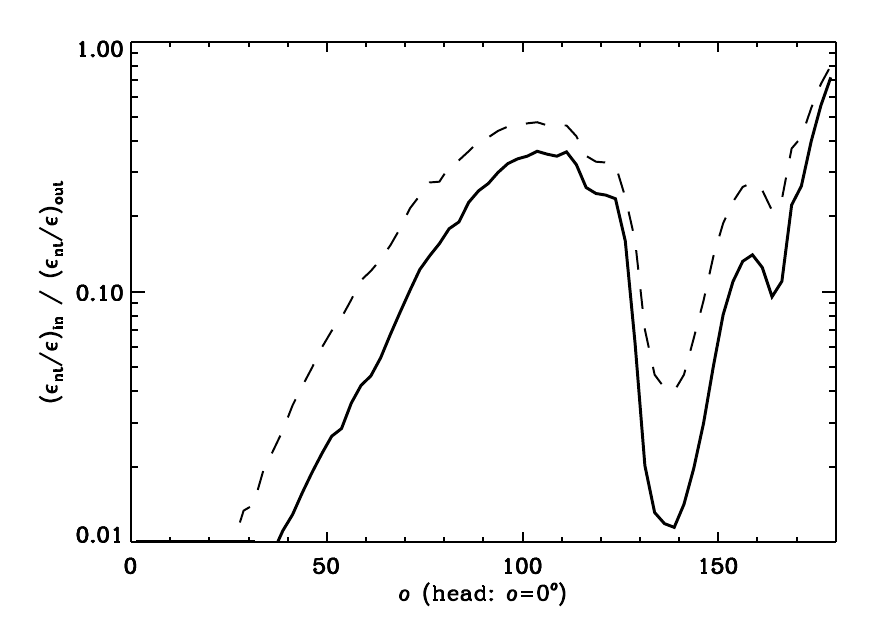}}
\caption{Fraction of the mean particle energy remaining after radiation losses along each streamline, assuming the conditions at periastron ($\phi=0$). The streamlines are ordered by angle $o$ (Fig.~\ref{fig:structure}). Dashed line shows the same at apastron ($\phi=0.5$).}
\label{fig:adiab}
\end{figure}

Our reference model for the emission has $\xi=1$ (acceleration timescale), $s=2$ (power-law slope of electron spectrum, no Maxwellian), and $B$ calculated using Eq.~\ref{eq:b1}. The parameters $\zeta_p$ and $\zeta_b$ are set to 1. At the apex of the termination shock at periastron ($d=0.098$\,AU), $\zeta_p=1$ corresponds to $\gamma_{\rm min}\approx 8\times 10^4$ and $\zeta_b=1$ corresponds to $B\approx40$\,G. Figure~\ref{fig:part_evolution}  illustrates the typical evolution of particles along two  streamlines, at periastron $\phi=0$. We label streamlines by the angle $o$ that their starting point makes with the binary axis, with $o=0\degr$ corresponding to the bow shock head on the binary axis,  $o=90\degr$ perpendicular to the binary axis, $130\degr \leq o\leq 180\degr$ to the back shock (see Fig.~\ref{fig:structure}, where the black dots identify the two  streamlines for which the particle evolution is shown). 

For the first streamline ($o=48\degr$), the initial particle distribution ranges from $\gamma_{\rm min}=6.8\times10^4$ to $\gamma_{\rm max}=8.8\times 10^6$ (top panels of Fig.~\ref{fig:part_evolution}). The highest energy electrons radiate away half of their energy on a scale $l\la 0.003$ (in units of the orbital separation, here $d=0.092$\,AU). For comparison, this is comparable to the spatial resolution at maximum grid refinement in our simulation. Properly resolving the cooling spatial scale within the RHD simulation would require an additional 1-2 levels of refinement, at large computational cost as mentioned in \S\ref{sec:total}. Even the lowest energy electrons cool on a small scale $l\la 0.05$ compared to the length of the streamline. Synchrotron and inverse Compton burnoff at high energy is seen in the evolution of the particle distribution (right panel). The late evolution of the distribution is set by adiabatic losses, which do not modify the shape of the distribution.

For the second streamline ($o=115\degr$), the initial distribution ranges from $\gamma_{\rm min}=1.2\times10^4$ to $\gamma_{\rm max}=3.1\times10^7$. The higher $\gamma_{\rm max}$ is due to the lower magnetic field at the shock (Eq.~\ref{eq:gmax}) while the lower $\gamma_{\rm min}$ is due to the lower pressure (Eq.~\ref{eq:gmin}). Radiative cooling is weakened by the distance and by the higher Lorentz factor (decreasing the comoving density of stellar photons). There is a moderate evolution of the particle distribution before the particles are re-energized by passing through the reflection shock. Compression at the shock heats the particles  and enhances the magnetic field. Radiative cooling is much more important in the subsequent evolution of the highest energy particles.

Without radiative losses, the particles lose $\approx 43$\% of their energy adiabatically within the box. The particle energy losses are enhanced by radiation. Figure~\ref{fig:adiab} shows the fraction of the total energy losses that are due to radiative losses depending on streamline. For each streamline, the ratio of the specific energy in non-thermal particles $\epsilon_{\rm nt}$ to the thermal energy $\epsilon$ is compared at the beginning and at the end of the streamline. 
\begin{equation}
\label{eq:ratio}
\frac{\epsilon_{\rm nt}}{\epsilon}\propto \frac{(\hat{\gamma}-1)}{p}{\int_{\gamma_{\rm min}}^{\gamma_{\rm max}} \gamma m_e c^2 \frac{dn}{d\gamma}d\gamma}.
\end{equation}
The value of  ${\epsilon_{\rm nt}}/{\epsilon}$ will be the same at the beginning and at the end of the streamline $({\epsilon_{\rm nt}}/{\epsilon})_{\rm in}=({\epsilon_{\rm nt}}/{\epsilon})_{\rm out}=\zeta_{p}$  if the particle losses (or gains) are only due to the adiabatic term {\em i.e.} the ratio $({\epsilon_{\rm nt}}/{\epsilon})_{\rm in} / ({\epsilon_{\rm nt}}/{\epsilon})_{\rm out}$ plotted in Figure~\ref{fig:adiab} will be 1 if there are no radiative losses. The figure shows this is not the case in our reference simulation: radiative losses dominate the overall energy losses as the particles propagate along the streamlines. The radiative losses are most important for the streamlines that start close to the apex ($o\la 45\degr$) and in the reflected shock region. Integrating $\epsilon_{\rm nt} \dot{N}$ over the whole flow, we find that $\approx 70\%$ of the power given to non-thermal particles is lost to radiation within our simulation box. Adiabatic losses have a minor influence on the spectrum and lightcurve: nearly identical results are obtained when our baseline calculation is run without taking adiabatic losses into account. 

We can only speculate on the feedback that radiative cooling could have on the flow dynamics, since our simulation does not take it into account. The shock region width is likely to decrease as the plasma loses pressure support, raising the density. Since the magnetic field intensity is tied to the density, this may cause particles to radiate even faster and at a higher synchrotron frequency than in our computation. Thin shell instabilities may also disrupt the interaction region. We leave this for future investigations.

\subsection{Emission maps\label{sec:maps}}
\begin{figure}
\resizebox{0.95\hsize}{!}{\includegraphics{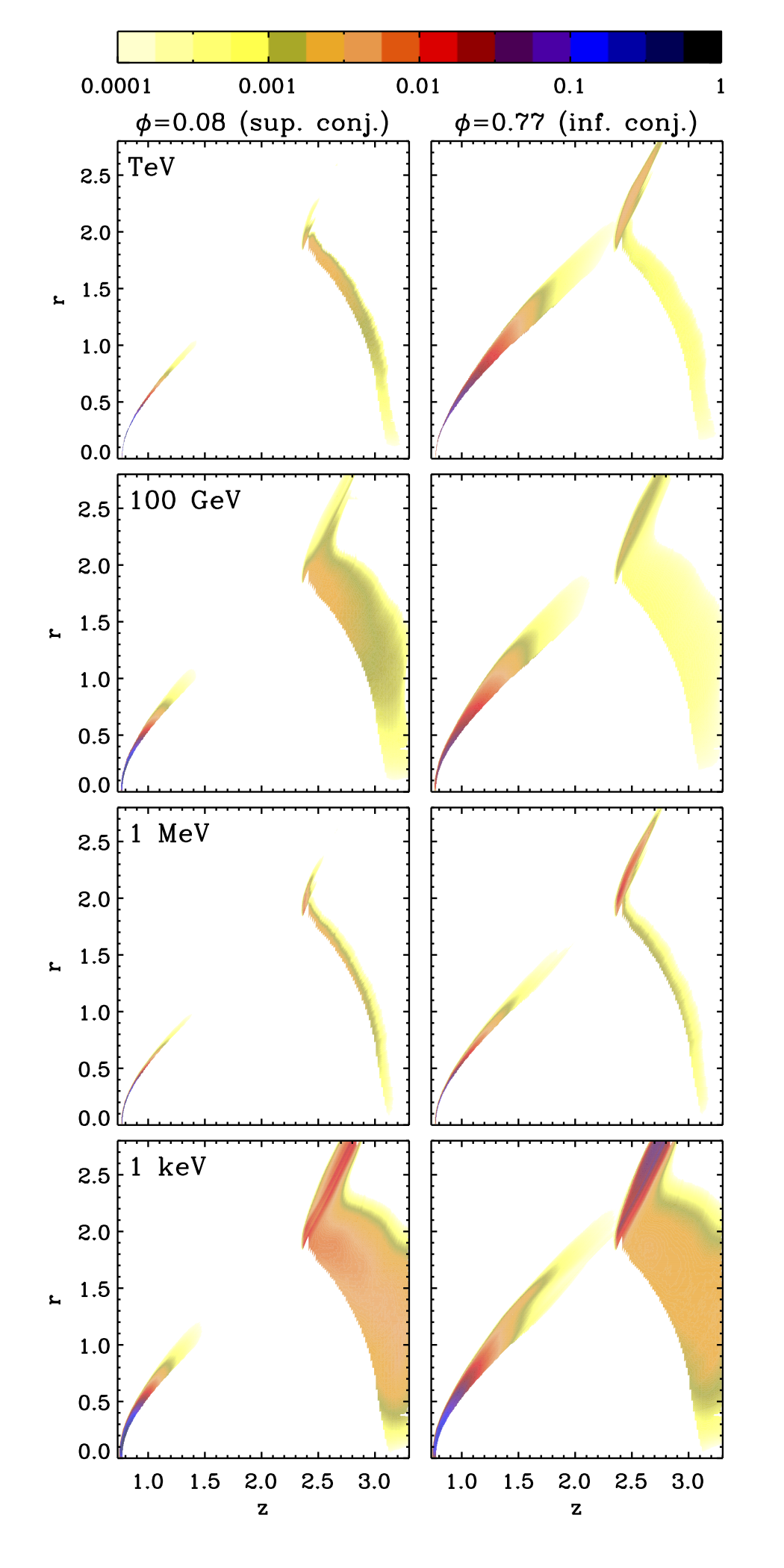}}
\caption{Emission maps of the shocked pulsar wind at various frequencies (top to bottom) and for orbital phases $\phi$ corresponding to the conjunctions (left and right columns). The emission is displayed on a logarithmic scale ranging from 1 down to $10^{-4}$, with 1 corresponding to the maximum value of the emission along the orbit at this frequency ({\em i.e.} each map is normalized to its maximum value over any pixel and any $\phi$). The VHE maps do not take the pair production opacity into account. The case shown here corresponds to $i=25\degr$ in Fig.~\ref{fig:incl}. The spatial scale is in units of the orbital separation $d$. See Figs.~A.1-A.4 in the online appendix for associated movies of the evolution with orbital phase.}
\label{fig:map}
\end{figure}

Emission maps were built for the baseline case with $i=25\degr$ (Fig.~\ref{fig:map}). The maps represent  the unabsorbed emissivity integrated over azimuth $\theta$ (Eq.~\ref{eq:flux}) $$\int \mathcal{D}_{\rm obs}^2  j \left({\nu}/{\mathcal{D_{\rm obs}}}\right) rd\theta.$$ This quantity was sampled along each streamline and binned to form maps at 1 keV, 1 MeV, 100 GeV, and 1 TeV. Figure~\ref{fig:map} compares the maps at superior ($\phi=0.08$) and inferior conjunctions ($\phi=0.77$). Animations showing the evolution at all orbital phases are available in the online appendix (Figs. A.1-A.4). Fast cooling concentrates emission at the highest frequencies to thin layers close to the pulsar termination shock (e.g. compare the synchrotron 1 MeV and 1 keV maps). The emission is more concentrated at $\phi=0.08$ than at   $\phi=0.77$  because the orbital separation is smaller ($d\approx 0.10$\,AU compared to $d\approx 0.15$\,AU), leading to stronger radiative losses.  The simulation box covers well the emission zones at 1 TeV and 1 MeV, but misses some of the 100 GeV and 1 keV, especially in the back region. Note, however, that the map flux scale is logarithmic so the impact on the overall lightcurve is negligible.

Reheating in the reflection shock region is easily seen in the maps, especially at 1 keV where a significant fraction of the flux may come from this region (and hence escape X-ray absorption by the stellar wind, see \citealt{2011MNRAS.411..193S}). The bow shock emission is concentrated towards the head while the back shock emission covers a much wider area. The VHE emission from the back region suffers less from $\gamma\gamma$ absorption and actually contributes nearly all the TeV flux when the opacity is highest (around $\phi=0.08$, see Fig.~\ref{fig:incl}).

\begin{figure}
\resizebox{0.95\hsize}{!}{\includegraphics{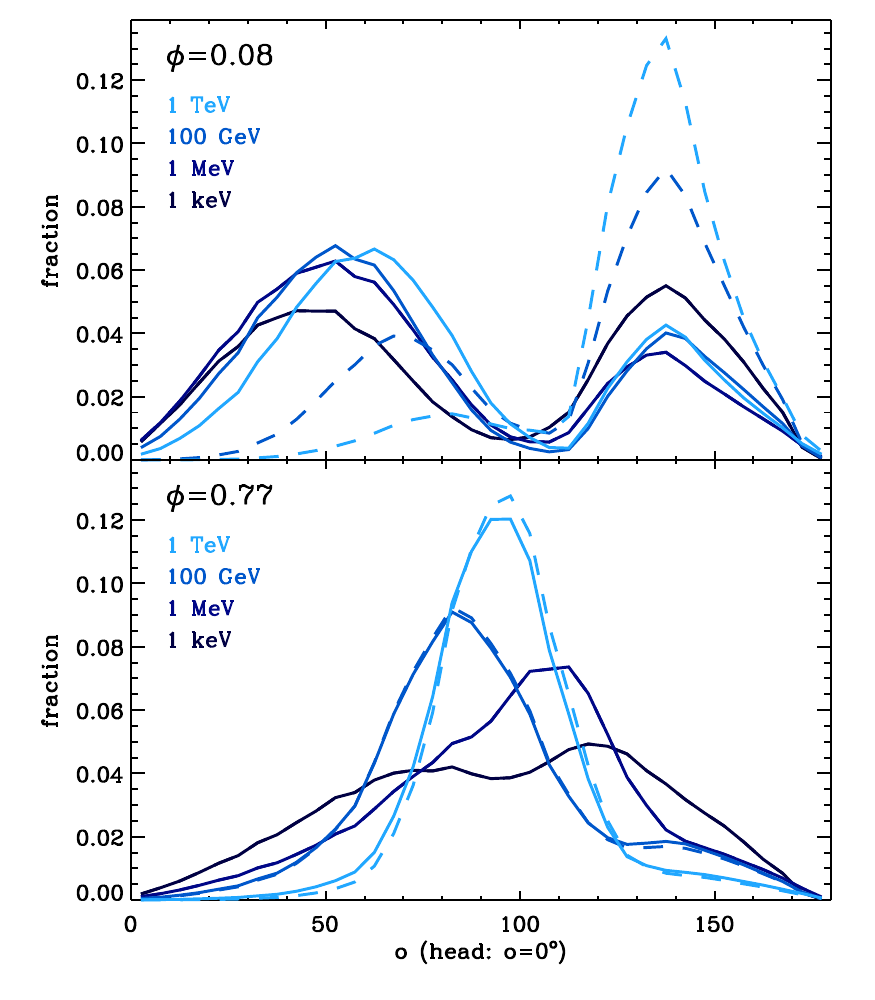}}
\caption{Contribution to the total flux by each streamline for our baseline model with $i=25\degr$. The streamlines are ordered by angle $o$ (Fig.~\ref{fig:structure}). Top panel is for orbital phase $\phi=0.08$ (superior conjunction), bottom panel for $\phi=0.77$ (inferior conjunction). The flux fractions in each panel correspond to the four frequencies mapped in Fig.~\ref{fig:map}. Solid lines correspond to the flux unabsorbed by pair production, dashed lines is for the absorbed flux (affecting  the fractions only at 100 GeV and 1 TeV).}
\label{fig:flux}	
\end{figure}

Figure~\ref{fig:flux} presents a different way of looking at where particles cool. We have plotted the integrated contribution of each streamline to the total emission at different frequencies. Streamlines that start at $o\leq 90\degr$ correspond to the head of the bow shock, streamlines with $o\geq 130\degr$ correspond to the back shock (Fig.~\ref{fig:structure}). The TeV inverse Compton emission originates mostly at streamline angles $o$ larger than for the 100 GeV emission, and the same applies when comparing the MeV and keV synchrotron emission. This is because the streamline initial magnetic field decreases with $o$, allowing for higher initial particle energies ($\gamma_{\rm max}$). The back shock contribution clearly dominates the absorbed VHE flux at $\phi=0.08$ (dashed lines). At $\phi=0.77$ the emission is much more concentrated in the streamlines with $o\approx 100\degr$, which pass through the reflected shock and strongly benefit from the relativistic Doppler boost since the flow in the back region is then aligned with the observer line-of-sight ($i=25\degr$). 

\subsection{Spectra and lightcurves\label{sec:spec}}
\begin{figure*}
\centerline{\includegraphics{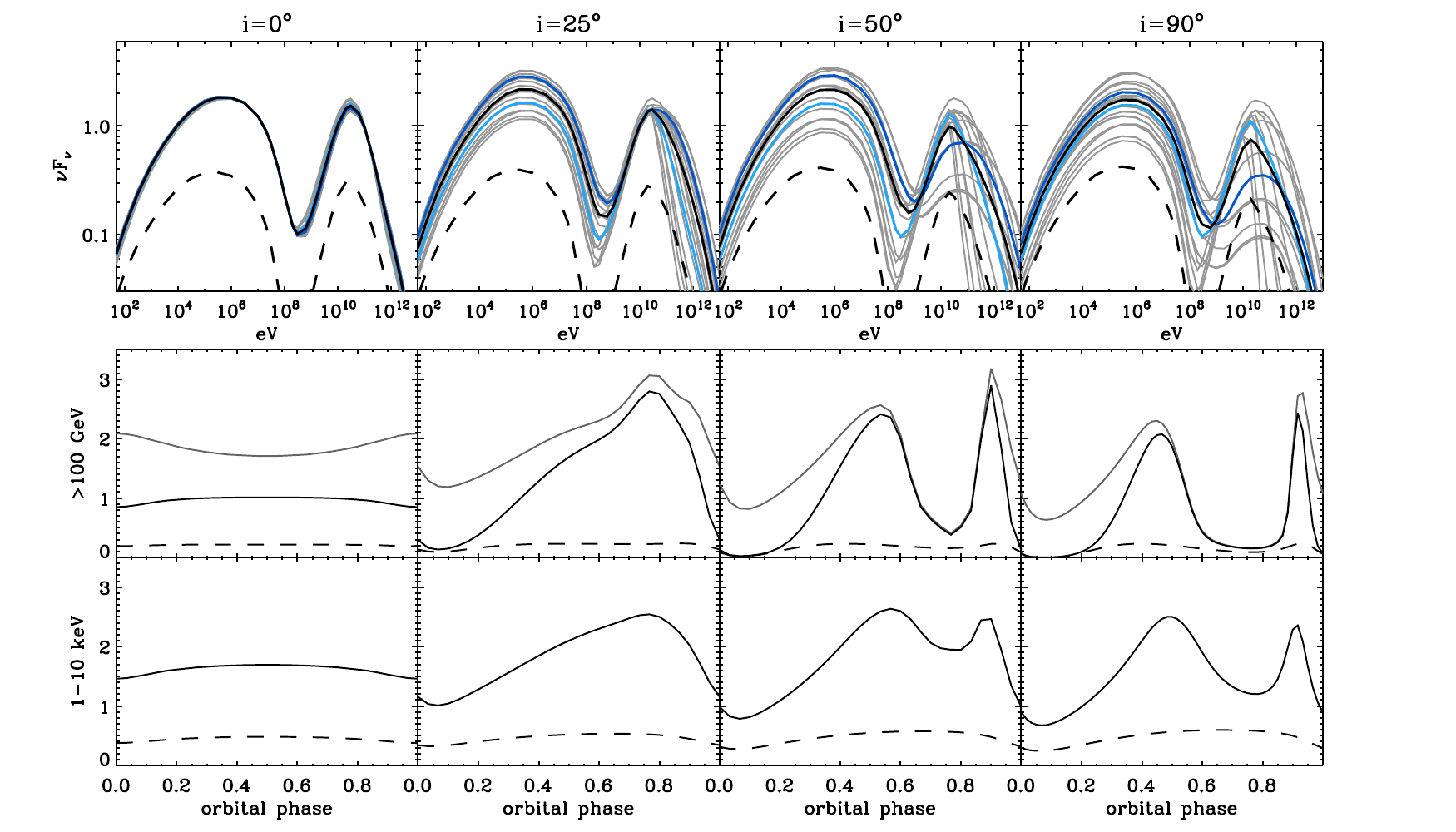}}
\caption{Spectral energy distributions and lightcurves depending on inclination for our reference model. Top panels: thick black line is the average spectrum, thick colored lines are the average INFC (dark blue) and SUPC (light blue) spectra (\S\ref{sec:spec}), thick dashed line shows the contribution of the back shock region to the average spectrum, thin grey lines show the spectral evolution with orbital phase. Bottom panels: VHE gamma-ray and X-ray lightcurves (thick solid lines). Again, the dashed line shows the back shock contribution. The thin grey line in the middle panels is the unabsorbed VHE flux.}
\label{fig:incl}
\end{figure*}Spectral energy distributions and lightcurves were computed for several system inclinations using our baseline model. The results are displayed in Fig.~\ref{fig:incl}. The spectra and lightcurves are normalised by a coefficient
\begin{equation}
{\cal K}=5\times 10^{-10}\, \left(\frac{2.5{\rm\,kpc}}{D}\right)^2 \left(\frac{\dot{E}_p}{7.6\times10^{35}\rm\,erg\,s^{-1}}\right)\rm\,erg\,cm^{-2}\,s^{-1},
\label{eq:norm}
\end{equation}
where we have explicited the dependence on the injected power in particles $\dot{E}_p=7.6\times10^{35}\rm\,erg\,s^{-1}$. $\dot{E}_p$ is related to the pulsar spindown power by $\dot{E}=\dot{E}_p (1+\sigma)$. The results can be scaled with $\dot{E}_p$, or $\sigma$, as long as $\dot{M}_w$ and $\dot{E}$ change in parallel to keep $\eta=0.1$ (\S2). Figure~\ref{fig:incl} shows the spectral energy distribution sampled at various orbital phases to highlight the spectral variability, the average spectral energy distribution (thick black line) and (in blue) the average spectrum corresponding to phases $0.45<\phi\leq 0.9$ (INFC) and $\phi\leq0.45$ or $\phi>0.9$ (SUPC), allowing a comparison with the H.E.S.S. spectral analysis in \citet{Aharonian:2006qwBIS}.

The size of the simulation domain limits how far we can follow particle cooling, as the maps of Fig.~\ref{fig:map} illustrate. The emission is thus necessarily incomplete below some energy, which we estimate to be $\la 100$\,eV (synchrotron emission component) and $\la 1$\,GeV (inverse Compton component) --- based on comparing spectra obtained with a reduced domain size.

The spectra produce broad band X-ray to TeV emission but, as could be expected (see \citealt{2013A&ARv..21...64D}), cannot reproduce the peaked GeV emission observed with the {\em Fermi}-LAT. This emission component requires a completely different population of electrons, with a narrow distribution in energy. \citet{2012arXiv1212.3222Z} speculated this could arise from the back shock but we find no obvious difference between bow and back shocks. The average spectrum from the back region is shown as a dashed line in the top panels. Emission from the back region can dominate near superior conjunction, when $\gamma\gamma$ absorption is important (see dashed lightcurve in the bottom panels), but its contribution to the average spectrum remains minor. The spectra of the bow and back region are similar ; they would need to have very different acceleration parameters to produce significantly different spectra  \citep{2012arXiv1212.3222Z,2014ApJ...790...18T}. We come back to the question of the origin of the HE gamma-ray emission in \S5.

We show lightcurves for the X-ray  (1-10\,keV) and VHE ($>100\rm\,GeV$) gamma-ray bands, where the spectra and orbital modulations are well-known from {\em Suzaku} and H.E.S.S. observations \citep{Takahashi:2008vu,Aharonian:2006qwBIS}. The lightcurves were computed by integrating $F(\nu)$ over the relevant energy range. When the system is seen face-on ($i=0\degr$), the VHE modulation is directly related to the varying stellar photon density which increases both inverse Compton emission and pair production. The synchrotron emission varies little. The synchrotron loss timescale increases at larger orbital separations since $\tau_{\rm sync}\sim b^{-3/2} \nu^{1/2}\sim d^{3/2}$ at a given frequency. However, the actual size of the computational domain increases as $d$, while the particle distribution slope $s=2$ ensures equal power per particle energy, so the emission should vary roughly as $d^{0.5}$, a  factor 1.4 from periastron to apastron, a bit more than what the full calculation gives (bottom left panel of Fig.~\ref{fig:incl}).

The emission received by the observer changes dramatically with the inclination angle of the system. The synchrotron emission is changed by relativistic boosting as ${\cal D}_{\rm obs}$ changes with orbital phase. The bow shock region creates a hollow cone of high-velocity material, surrounding a filled cone of lower-velocity material flowing away from the back shock. Increasing the inclination boosts the X-ray emission around inferior conjunction $\phi_{\rm inf}\approx 0.77$, when the shocked flow is oriented towards the observer, and de-boosts the X-ray emission around superior conjunction $\phi_{\rm sup}\approx 0.08$, when the flow is directed away. At higher inclination the observer line-of-sight starts crossing the emission cone of the bow shock along its full length. This result in maximum boost at $\phi_{\rm inf}$ and $40\degr\la i\la 50\degr$ when the shocked flow is going directly in the direction of the observer. For $i\ga50\degr$ the line-of-sight crosses first one edge of the hollow cone then the other, resulting in double-peaked emission with a minimum at $\phi_{\rm inf}$. At $i=90\degr$, the line-of-sight crosses the cone first at orbital phases $0.54\la \phi\la 0.60$ and  then at $0.87\la \phi\la 0.89$, in agreement with the position of the two peaks. Although the cone is symmetric, the  second peak is narrower because of the faster orbital motion during the second crossing (nearer to periastron passage).  

\begin{figure}
\resizebox{0.95\hsize}{!}{\includegraphics{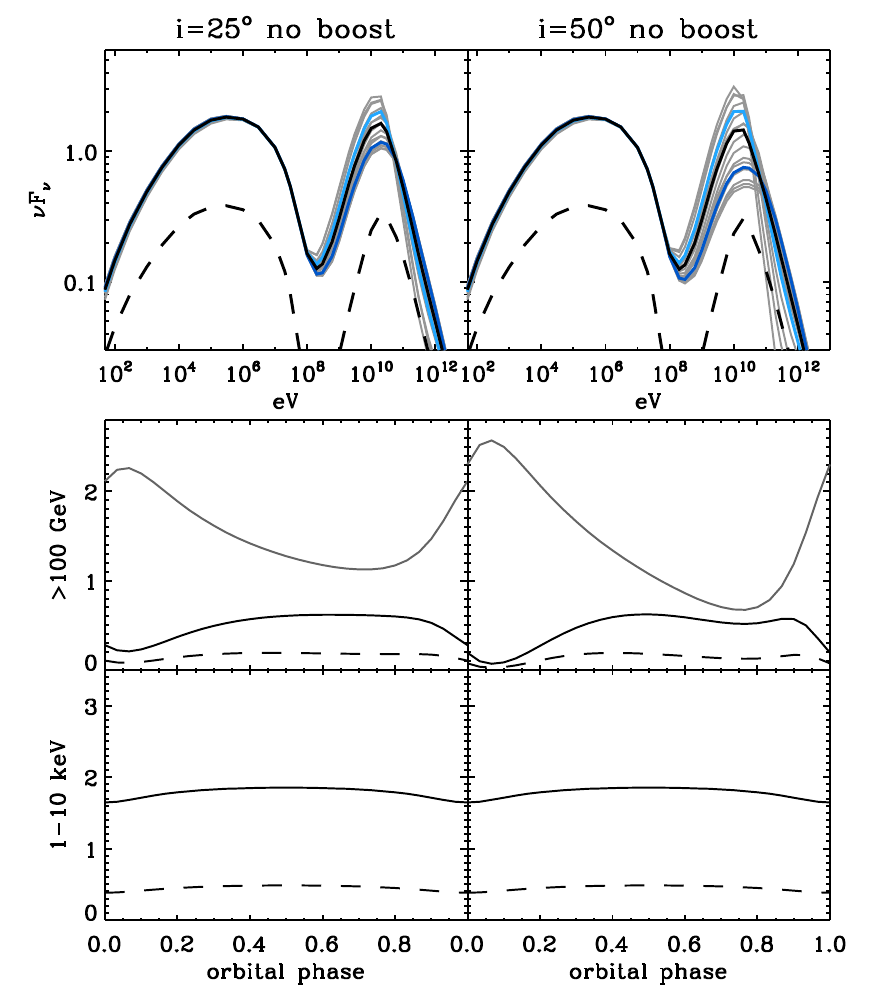}}
\caption{Same as Fig.~\ref{fig:incl} for $i=25\degr$ and $50\degr$ except that relativistic aberrations has not been taken into account when computing the emission.}
\label{fig:relat}
\end{figure}

The VHE emission is influenced by the anisotropy of the inverse Compton and pair production cross-sections. Inverse Compton emission is enhanced  around $\phi_{\rm sup}$, when stellar photons are backscattered towards the observer, and diminished around $\phi_{\rm inf}$, when the stellar photons are forward-scattered (Fig.~\ref{fig:relat}). For the same reasons, the pair production opacity is important around $\phi_{\rm sup}$ and lower at $\phi_{\rm inf}$. The latter can be verified by comparing the dark (absorbed) and grey  (unabsorbed) lines in the middle panels of Fig.~\ref{fig:incl}. However, the effects of Doppler boosting  dominate the VHE lightcurve. Figure~\ref{fig:relat} shows the expected lightcurve at $i=25\degr$ but with the relativistic effects turned off when computing the emission (the electron populations are identical). The X-ray synchrotron lightcurve is nearly constant in the absence of Doppler boosting effects. Comparing with the full calculation (Fig.~\ref{fig:incl}), relativistic effects displace the peak VHE and X-ray emission towards $\phi_{\rm inf}$ and then lead to a double-peaked structure at higher $i$.  Relativistic boosting also hardens the VHE spectrum around $\phi_{\rm inf}$. The intrinsic anisotropic inverse Compton emission is harder at these orbital phases because  scattering is increasingly within the Thomson regime when the stellar photons are closer to being forward-scattered \citep{Dubus:2007oq}. This effect is amplified by the bulk Doppler boosting. The strong spectral evolution with orbital phase at high $i$ can be followed in the top panels of Fig.~\ref{fig:incl}. The grey lines show the spectral energy distribution at different phases. The inverse Compton spectrum pivots around 100 GeV for $i=50\degr$. Above this energy, maximum emission occurs around $\phi_{\rm inf}$ (the INFC spectrum is brighter) whereas, below this pivot energy, the gamma-ray emission peaks around $\phi_{\rm sup}$ (the SUPC spectrum is brighter). The inverse Compton lightcurve at different frequencies can thus behave in antiphase because of the subtle hardening effects   brought about by scattering angle and bulk Doppler boost.

\subsection{Exploring parameter space\label{sec:params}}
\begin{figure*}
\centerline{\includegraphics{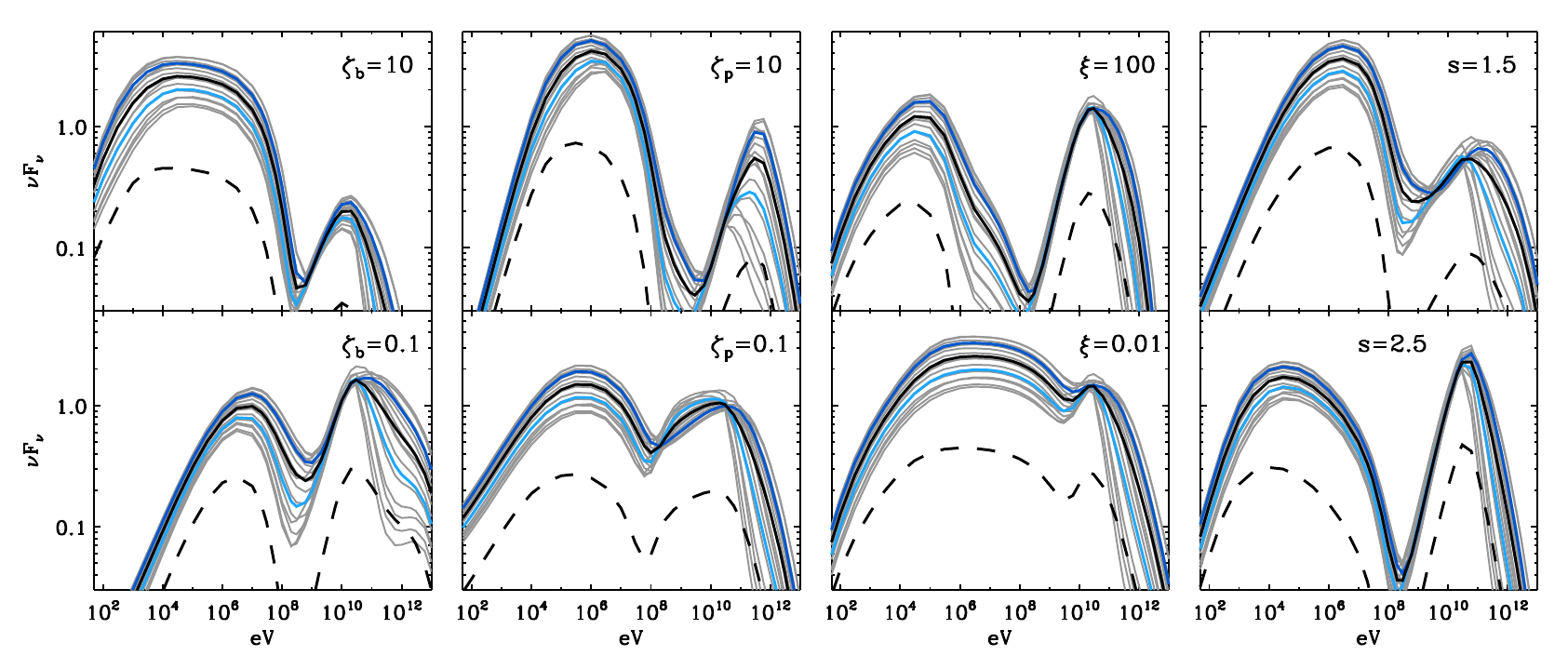}}
\caption{Dependence of the spectrum on the model parameters. The spectra should be compared to the baseline model with $\zeta_b=\zeta_p=\xi=1$, $s=2$ and $i=25\degr$ (second column of Fig.~\ref{fig:incl}). For line labels, see caption to Fig.~\ref{fig:incl}. }
\label{fig:param}
\end{figure*}
We carrried out a limited exploration of the parameter space around our reference model. The dependence on inclination $i$ has already been shown in Fig.~\ref{fig:incl}. The dependence on the other parameters, namely  $\zeta_{p}$, $\zeta_{b}$, $\xi$, and $s$, is shown in Fig.~\ref{fig:param}. The normalisation of the spectra is the same as for our reference model (Eq.~\ref{eq:norm}). We remind that $\zeta_{p}$ controls the mean energy of the particles (Eq.~\ref{eq:gmin}), $\zeta_{b}$ controls the magnetic field intensity at the shock (Eq.~\ref{eq:b1}), $\xi$ controls the maximum particle energy at the shock (Eq.~\ref{eq:bohm}), and $s$ is the slope of the injected power-law distribution of electrons.

We successively changed the value of each parameter, keeping the others to their reference value. A higher $\zeta_b$ increases synchrotron losses, leading to a pronounced $\nu F_\nu \sim \nu^{(2-s)/2}\sim \nu^0$ spectrum of cooled particles and lowering the inverse Compton emission component. Conversely, a lower $\zeta_b$ increases the inverse Compton component relative to the synchrotron component and, in our case, leads to a hard synchrotron spectrum because the inverse Compton energy losses are in the Klein-Nishina regime. A higher $\zeta_p$ increases $\gamma_{\rm min}$ and thus narrows down the energy range of the injected power law. The low-energy slope of the synchrotron component for $\zeta_p=10$ corresponds to the $\nu F_\nu \sim \nu^{4/3}$ expected for a tail of emission from electrons at  a high  $\gamma_{\rm min}$ whereas, in the $\zeta_p=0.1$ case, the slope is the  $\nu F_\nu \sim \nu^{(3-s)/2}\sim \nu^{1/2}$ slope expected from uncooled electrons emitting in our frequency range with (smaller) Lorentz factors $\geq \gamma_{\rm min}$. Changing the acceleration timescale $\xi$ directly impacts the maximum synchrotron frequency, but has no influence on the inverse Compton emission because, in our case, $\gamma_{\rm max}$ is always high enough for the interaction with stellar photons to occur in the inefficient Klein-Nishina regime. Finally, changing the slope $s$ of the injected distribution, unsurprisingly, produces a harder synchrotron spectrum when the electron distribution is harder (smaller $s$). The ratio of synchrotron to inverse Compton emission is also higher because a harder distribution implies more very high energy electrons that radiate more efficiently synchrotron emission compared to inverse Compton emission in the Klein-Nishina regime.

We do not show how the TeV and X-ray modulations were affected by the changes in parameters in Fig.~\ref{fig:param}. The reason is that the modulations did not change much compared to the reference model. These lightcurves are predominantly shaped  by the inclination { rather than by changes in the other parameters}. All cases are also comparably radiatively-efficient: the bolometric luminosities of the different models vary, in normalised units { (Eq.~\ref{eq:norm})}, between $\approx$ 13 ($\xi$=100) and $\approx 35$ ($\xi=0.01$). As in the reference case, most of the pulsar power is converted into radiation.

\section{Application to LS 5039}
These results guided us towards a model reproducing the emission from LS 5039 based on our RHD simulation. This model is compared against the observed spectral energy distribution of LS 5039 in Fig.~\ref{fig:ls5039spec} and the X-ray, MeV, GeV, and TeV lightcurves in Fig.~\ref{fig:ls5039light}. The cost of the calculations (several hours per model) does not allow an extensive exploration of parameter space. The parameter combination given here is only indicative of what seems to work {\em i.e.} this is not a best-fit model. 

\subsection{Model parameters}
The main drivers in deriving this model were (1) reproducing the VHE spectral variations ; (2) accounting for the comparable X-ray and VHE gamma-ray fluxes ; (3) understanding the origin of the HE gamma-ray emission. We start with the latter.

As the results from the previous section should make clear, the HE gamma-ray emission observed with the {\em Fermi}-LAT requires an additional component. In principle, a low value of $\xi$ pushing the synchrotron component to GeV energies might account for the {\em Fermi}-LAT spectrum with an exponential cutoff (at the price of supposing a faster-than-Bohm acceleration timescale). However, the HE modulation would then be in phase with the X-ray modulation, which is ruled out by the observations. The HE modulation is actually consistent with expectations for inverse Compton scattering of stellar photons \citep{2009ApJ...706L..56ABIS}. 

\begin{figure}
\resizebox{0.95\hsize}{!}{\includegraphics{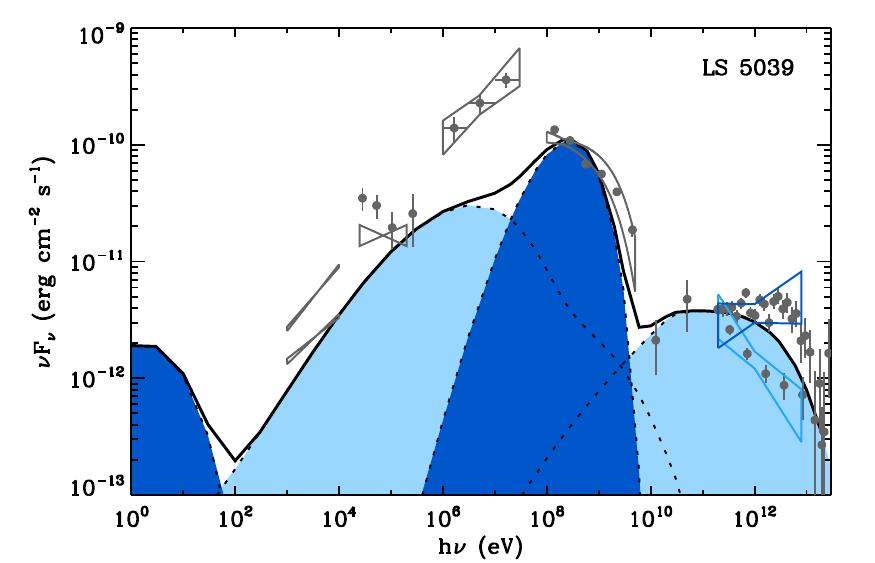}}
\resizebox{0.95\hsize}{!}{\includegraphics{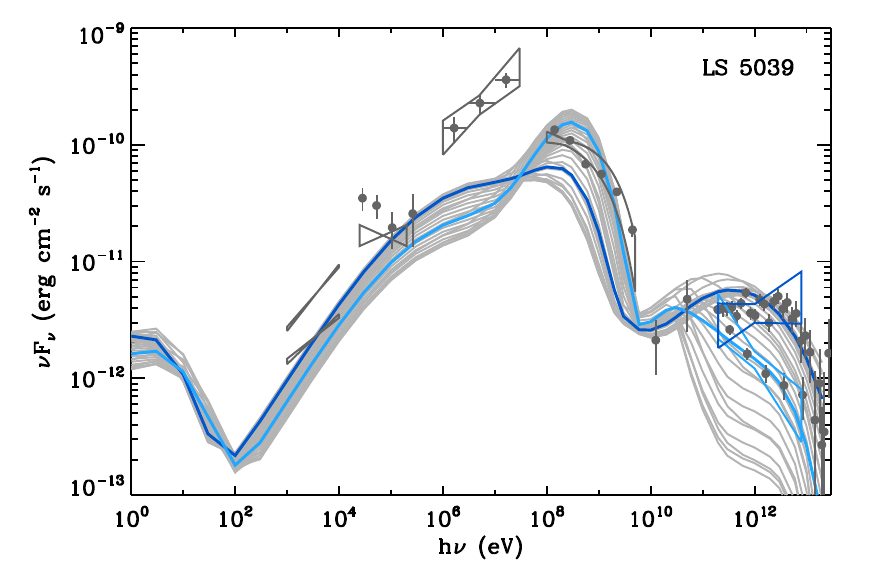}}
\caption{Our best-adjusted model to the observed spectral energy distribution of LS 5039. Top: the black curve is the average model spectrum ; the synchrotron ($\sim 1$\,eV) and inverse Compton ($\sim 1$\,GeV) contributions from the Maxwellian population of electrons are in dark blue ; the contributions from the power-law population are in light blue. Bottom: the thick, dark blue curve (resp. light blue curve) represents the INFC (resp. SUPC) spectrum, the thin grey curves represent the evolution of the SED in orbital phase steps of 1/30. From left to right, the data shown are: the orbital maximum and minimum X-ray bowties from {\em Suzaku} \citep{Takahashi:2008vu}, the  BATSE data points and INTEGRAL bowtie \citep{Harmon:2004ml,Hoffmann:2008ys}, the average COMPTEL data points with bowtie \citep{2014A&A...565A..38C}, the average {\em Fermi}-LAT spectrum with data points from 100 MeV to 50 GeV and the best-fit power-law with exponential cutoff \citep{2012ApJ...749...54H}, the H.E.S.S. INFC (dark blue bowtie) and SUPC spectra (light blue bowtie) with the associated data points from 100 GeV to 30 TeV \citep{Aharonian:2006qwBIS}.}
\label{fig:ls5039spec}
\end{figure}

We explored the possibility that the HE emission could be due to the inverse Compton emission from a narrow Maxwellian distribution of electrons, as we had for the case of PSR B1259-63 in \citealt{2013A&A...557A.127D}. We assumed that a fraction of the pulsar wind particles injected at the shock  are accelerated to a power law, accounting for the broad band X-ray to TeV emission, while the rest are randomized to this Maxwellian distribution. Adjusting the HE spectrum with the inverse Compton emission from the Maxwellian fixes its mean Lorentz factor to  $\gamma_{t}\approx 5000$. This also fixes $\zeta_{p}$ to $\approx 0.017$ through Eq.~\ref{eq:gmin}. The available specific internal energy is {\em a priori} identical for the Maxwellian and power-law distributions so $\zeta_{p}$ is thus fixed for both populations of electrons. The relative contribution of each population is adjusted by the fraction of the particle density going to each (or, equivalently, total energy since $\epsilon_{\rm nt}$ is the same). The contributions to the average spectrum from each population of particles are highlighted in the top panel of  Fig.~\ref{fig:ls5039spec}. 
 
The H.E.S.S. INFC spectrum is hard, best described as a power law of photon index of 1.8 combined with an exponential cutoff at 8.7 TeV. The SUPC spectrum is a steeper power law with an index of 2.5 (Fig.~\ref{fig:ls5039spec}). The two spectra pivot at an energy $\approx 200$\,GeV. Reproducing both the hard INFC spectrum and the comparable levels of X-ray and VHE emission turned out to be difficult.  The models explored in \S4 all cut off around 100 GeV because the high-energy particles responsible for this emission are strongly cooled by synchrotron losses. Lowering $\zeta_{b}$ decreases the mean magnetic field, hence increases the VHE cutoff, but also lowers the X-ray flux relative to the VHE gamma-ray flux. There is a trade-off between having enough synchrotron emission to account for the X-ray flux, and lowering $\zeta_{b}$ to enable the highest-energy particles to radiate enough VHE photons. As discussed in \S\ref{sec:params}, changing $\xi$ has little influence on the VHE spectrum so we kept $\xi=1$. Limited exploration showed the best agreement was obtained by slightly lowering $\zeta_{b}$ to 0.5 and by taking an injection slope $s=1.5$ instead of 2. 

The inclination has an effect on both lightcurves and spectral variations. The VHE spectral variations are more pronounced with higher $i$ although too high an inclination results in a INFC spectrum with a flux that is too low  (Fig.~\ref{fig:incl}). A high inclination also results in a pronounced dip of emission at $\phi_{\rm inf}=0.77$. The H.E.S.S. lightcurve appears double-peaked with a shallow minimum at $\phi_{\rm inf}$, favoring a model with $25\degr < i < 50\degr$ (Fig.~\ref{fig:incl}). The X-ray modulation is single-peaked at $\phi_{\rm inf}$, favouring models towards the low end of this range of $i$. Using $i=35\degr$ turned out to be a good compromise.

The ``best-adjusted'' model shown in Fig.~\ref{fig:ls5039spec}-\ref{fig:ls5039light} has $\zeta_{p}\approx 0.017$, $\zeta_{b}=0.5$, $\xi=1$, $s=1.5$, $i=35\degr$ and an injected population of particles consisting of a Maxwellian plus a power law. An additional parameter is the value of $\eta=0.1$ that we fixed all throughout this study. The hard injection spectral slope $s=1.5$ hints at reconnection rather than Fermi acceleration.  Adjusting the model to the observations requires that the injected power in particles is  $\dot{E}_p\approx 10^{35}$ erg\,s$^{-1}$. The majority of the particles or available power (88\%) goes to the Maxwellian at the termination shock. Only 12\% goes to the particles distributed as a power law. However, these power-law particles are more radiatively efficient than those in the Maxwellian: about 80\% of the power injected as a power law ends up radiated away within the simulation domain (like the models shown in \S4) compared with only 25\% of the power injected as a Maxwellian. Hence, most of the radiation comes from the injected power law but most of the power is in the Maxwellian. As a consequence, most of the particles evolve adiabatically in this model. 

\subsection{Comparison to the observations}
\begin{figure}
\resizebox{\hsize}{!}{\includegraphics{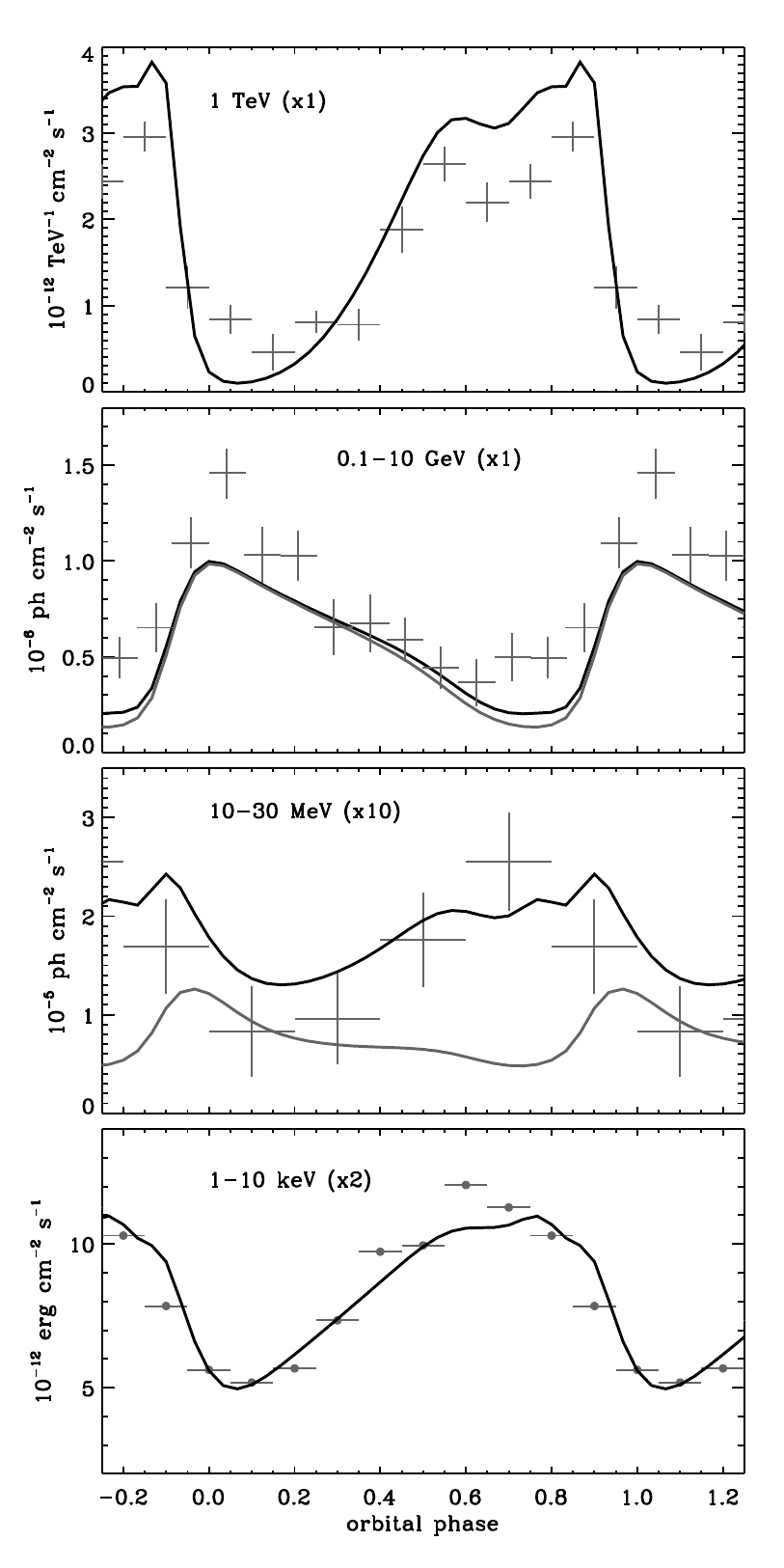}}
\caption{Comparison between the LS 5039 model (Fig.~\ref{fig:ls5039spec}) and the observed lightcurves. The model 10-30 MeV and 1-10 keV lightcurves were multiplied by a factor 10 and 2, respectively, to match the observations in \citet{Takahashi:2008vu,2014A&A...565A..38C}. The grey lightcurve n the 0.1-10 GeV and 10-30 MeV panels represents the contribution from the relativistic Maxwellian component. The VHE and HE gamma-ray lightcurves are taken from \citet{Aharonian:2006qwBIS,2009ApJ...706L..56ABIS}, respectively.}
\label{fig:ls5039light}
\end{figure}

The (rough) adjustment provides a reasonable, albeit imperfect, description of the data. The X-ray flux is too low by a factor $\sim 2$ in the present model. X-ray emission arising beyond our computational domain might account for this mismatch (\S4.4 and Fig.~\ref{fig:map}). A higher $\zeta_b$ would raise the X-ray flux, but lower the VHE cutoff in the INFC spectrum to values that would not be consistent with the observations. The largest discrepancy is the COMPTEL data at MeV energies, recently associated with LS 5039 \citep{2014A&A...565A..38C}. Although the spectral slope of the model, as well as its evolution with orbital phase (Fig.~\ref{fig:ls5039light}), are compatible with the observations (the COMPTEL spectrum is harder at SUPC phases), the flux is clearly underestimated by an order-of-magnitude. A higher $\zeta_b$ would raise the synchrotron luminosity but a cooling break in the synchrotron spectrum is hard to avoid and this would also be at the expense of the hard VHE INFC spectrum. If $\zeta_{b}$ is low then the synchrotron spectrum can extend from X-ray to MeV, as in \citet{Takahashi:2008vu} where adiabatic cooling is assumed to dominate over radiative cooling, but our model shows the inverse Compton component would then be too narrow and luminous (see Fig.~\ref{fig:param}).  Our model is unlikely to be able to account for the observed MeV flux without additional ingredients (discussed in \S6), unless the flux is contaminated by diffuse emission or the flux from other sources due to the poor angular resolution at these energies. Progress is much needed in this difficult observational band.

The orbital modulations from the model are compared with the observations in Figure~\ref{fig:ls5039light}. The fluxes were calculated in each band in the units given by the observations. Because of the mismatch in X-ray and MeV fluxes, we had to multiply the model flux by a factor 2 and 10 (respectively) to obtain a level comparable to the observations. The lightcurves are reasonably well reproduced. A slightly larger inclination would deepen the VHE minimum at $\phi\approx 0.7$. The Maxwellian component dominates in the HE band, with a modulation in anti-phase with the other wavelengths. The simulation output shows that the HE emission is concentrated at the head of the bow shock and is not affected much by relativistic boosting. The HE modulation is dominated by the variation with phase of the stellar photon density and scattering angle, resulting in peak emission at $\approx \phi_{\rm sup}$. The Maxwellian component also contributes some flux in the 10-30 MeV range. A small peak at $\phi\approx 0.75$ is visible in the TeV, MeV, and X-ray lightcurves. This appears to be due to the observer line-of-sight grazing the top of the emission cone at this orbital phase when $i=35\degr$. The small-scale, stable, flaring structures observed in the X-ray modulation lightcurve \citep{2009ApJ...697L...1K} might thus be directly related to small structures in the shocked flow that are probed when our line-of-sight passes through. Emission from the back shock region provides some residual TeV flux around $\phi_{\rm sup}$, when emission from the bow shock region is strongly absorbed by pair production. This is still insufficient to explain the VHE detections. Emission from the cascade, initiated when the newly-created $e^+e^-$ pairs are able to radiate VHE gamma rays, is very likely to be responsible for the residual flux at  $\phi_{\rm sup}$. \citet{2010A&A...519A..81C} found from their study of cascade emission that a good fit required an inclination $i\approx 40\degr$, consistent with the present model.

The synchrotron emission from the Maxwellian component peaks in the visible band, where it will be difficult to detect against the bright $V=11.2$ companion O star. This emission is modulated, with a lightcurve shape (not shown) similar to the MeV modulation. The peak-to-peak amplitude of the $V$ band modulation is 0.25 mJy. This translates into a 1.3 mmag modulation, below the current upper limit of 2 mmag \citep{2011MNRAS.411.1293S}.

\section{Discussion}

\subsection{Influence of the hydrodynamics on the flow emission}
Our motivation for developing this radiative code, based on a relativistic hydrodynamical simulation, was to obtain a more realistic and coherent treatment of the emission geometry, adiabatic losses, and Doppler boosting. We discuss these points in turn below.

The simulation shows a complex shock structure, fully containing the pulsar wind, with a reflected shock that re-energizes the shocked pulsar wind. In our models, most of the highest-energy emission remains concentrated towards the head of the bow shock where the electrons cool quickly.  Hence, both the spectral energy distribution and the modulations are predominantly shaped by the head of the bow shock region.  This is fortunate as the back shock and reflected shock structures are likely to have some dependence on our choice of Mach number and could change in the presence of orbital motion, strong mixing, dynamically important radiative losses or magnetic fields, etc. It is primarily at lower frequencies, notably in X-rays, that these structures contribute significantly, when the particles cool more slowly (on larger spatial scales) and/or re-heated to mild energies. These conclusions depend on the assumptions that we made on what happens at the various shocks, namely that there is no  difference in particle acceleration between the bow and back shock, and that the reflected shock only compresses particles. The first assumption is likely to be incorrect at some level because pulsar winds are not isotropic. Some latitude dependence is expected in the pulsar wind due, for instance, to differences between the propagation of the high-latitude regions and the equatorial region (defined by the pulsar rotation axis) where the pulsar wind is striped and prone to reconnection. This may be manifest in a latitude-dependence of the Lorentz factor of the pulsar wind, as seems to be required by models of the Crab pulsar wind nebula \citep{Bogovalov:2002wo,2014MNRAS.438..278P}, and/or a dependence of the pulsar wind magnetisation $\sigma$ with distance \citep{2012arXiv1212.3222Z,2014ApJ...790...18T}. A latitude-dependent $\epsilon$ or $B$ could have important observational signatures, even if there is no dramatic change in the structure of the flow \citep{Vigelius:2007vm,2012MNRAS.419.3426B}. On the second assumption, having the reflected shock only adiabatically compress the particles has some justification because it is usually found to be difficult to accelerate particles at shock in pair plasmas except in special circumstances (very low magnetisation, shock-induced reconnection of the striped wind). Yet, we cannot exclude that particles are re-accelerated to a power law, with some influence on the overall emission. We leave the exploration of these possibilities to future work.

The models we have explored are radiatively very efficient despite the fast flow timescale, the size of the emission region, and the decreasing magnetic field strength with distance. Adiabatic losses play a minor role in the high-energy emission, excluding them as the main driver of the X-ray and VHE modulation in LS 5039 as proposed by \citet{Takahashi:2008vu}. In principle, strong radiative losses should be taken into account in the dynamics of the flow region. These would result in a narrower and denser shocked flow, experiencing a higher magnetic field (\S\ref{sec:cooling}). However, in order to reproduce the spectral component seen with the {\em Fermi}-LAT, we have proposed that most of the particles in the shocked flow are actually injected in the form of a Maxwellian component rather than accelerated to a power-law distribution.  If the particles in the Maxwellian were only randomized at the shock, their mean Lorentz factor $\gamma_t$ corresponds to the Lorentz factor $\Gamma_p$ of the pulsar wind so $\Gamma_p\approx \gamma_t=5000$. These particles do not cool efficiently in the shocked flow and actually dominate the energy budget\footnote{We have not taken into account the inverse Compton emission from the particles in the free pulsar wind, which would contribute to the flux in the {\em Fermi}-LAT band much like the Maxwellian \citep{2014ApJ...790...18T}. Less than 50\% of the energy is lost to radiation if $\Gamma_p=5000$ according to Fig. 4 of \citet{2008A&A...488...37C}. This is an upper limit since their geometry did not include the back shock, hence the pulsar wind was free to propagate to infinity.}. Hence, perhaps counter-intuitively, the flow remains essentially adiabatic and the assumption of the simulation is verified.

Doppler boosting has a very strong effect in shaping the modulation lightcurves. The geometry is basically a rotating cone whose emission is boosted at the phases where its wings cross the observer line-of-sight. The result is a  double-peaked modulation at high inclinations, affecting the synchrotron emission and the VHE inverse Compton emission. In our models, the modulation due to the anisotropy of the inverse Compton cross-section is more important than the Doppler modulation only at lower energies, when the scattering occurs in the Thomson regime. The dependence of the lightcurve shape on inclination allows us to constrain $i$ to $\approx 35\degr$.  A side effect of Doppler boosting is that it contributes to steepening the VHE emission near superior conjunction, erasing the dip around a few 100 GeV in the SUPC spectrum due to $\gamma\gamma$ absorption and typically seen in previous models \citep[e.g.][]{Dubus:2007oq,Khangulyan:2007me,2012ApJ...761..146Y}. Our simulation assumed a pulsar wind Lorentz factor $\Gamma_p=7$, not quite high enough to obtain ultrarelativistic shock conditions along most of the bow shock. A higher $\Gamma_p$ could enhance the contribution from the wings, although we consider this unlikely given the sharp decrease in flux as the shock becomes mostly transverse (Fig.~\ref{fig:map}). Even if a simulation with a higher $\Gamma_p$ is desirable, we do not expect our results to change much.

\subsection{Towards more realistic models}
Adjusting to the observations of LS~5039 highlights both the difficulties and the progress to be expected from the present approach. Parameters that would be well-suited to the observations in one energy band are easily discarded as a result of the comprehensive approach taken here, because they fail to reproduce the observations in another band. 

We have discussed in \S5 the difficulties in reproducing both the level of the X-ray flux and the hard VHE spectrum. Part of the difficulty may be alleviated if cascade emission is taken into account. It would contribute to the VHE flux at all phases, not only at superior conjunction, and also to the X-ray emission via synchrotron radiation from the pairs  \citep{Bednarek:2007qd,Bosch-Ramon:2008vd,2010A&A...519A..81C}. Combining the present model with a 3D cascade model represents a daunting task. 

The COMPTEL flux level presents a similar challenge to models. The modulation, in phase with the X-rays and in anti-phase with the HE gamma rays, excludes that the COMPTEL emission arises from the same electrons responsible for the HE gamma-ray emission. It is more natural to attribute it to the extension of the synchrotron emission \citep{Takahashi:2008vu}, yet it appears very difficult to achieve this without a high magnetic field. As explained above in \S6.1, more complex dependencies of $\epsilon$ or $B$, motivated by the physics of pulsar winds and their termination shock, might be able to simultaneously explain the VHE emission and the strong synchrotron emission. 

Our results are based on a single simulation with a given $\eta$. They should hold qualitatively for other values of $\eta$. The strongest impact would certainly be on the value of the inclination required to adjust the observations since the cone opening angle depends directly on $\eta$. More subtle effects may appear if $\eta$ changes along the orbit. This is to be expected at some level in LS 5039 because the stellar wind is still in its acceleration phase when it encounters the pulsar wind at a distance of one stellar radius from the surface of the star. Using a beta law for the stellar wind velocity, we find that this leads to relative changes of $20\%$ in $\eta$. The opening angle of the cone would be slightly higher at periastron than at apastron. Much stronger effects are expected in the case of gamma-ray binaries like \object{LS I+61\degr303} and \object{PSR B1259-63} where the pulsar wind interacts with a dense equatorial outflow from the companion star. Modelling these systems requires orbit-dependent simulations. Besides the orbital phase dependency of $\eta$, such simulations will also be able to address the impact of orbital motion on the shape of the interaction cone. We expect that the leading arm (the part of the shocked pulsar wind that moves into the stellar wind due to orbital motion) will be compressed and the trailing arm will expand (see \citealt{2012A&A...546A..60L}, \citealt{2014arXiv1411.7892B} and references therein). The impact on the lightcurves should be limited since the emission arises mostly from the innermost, less-affected regions. The back shock in our simulation looks similar to the ``Coriolis shock" identified by \citet{2011A&A...535A..20B} and \citet{2012A&A...544A..59B,2014arXiv1411.7892B} in simulations including orbital motion. We speculate that the presence of a back shock is not related to the Coriolis force, and note that some of our previous simulations and those of \citet{2008MNRAS.387...63B} indeed show full confinement without orbital motion, albeit with a different back shock geometry  (\S4.1). Dedicated 2.5D (cylindrical) relativistic simulations would be useful to clearly define the conditions for full confinement, the shape of the structure, and resolve the issue.

The parameter $\zeta_{b}$ imposes the value of the magnetic field at the apex of the bow shock: $B\approx 20\,$G at a distance $\approx 3\times10^{11}\,$cm from the pulsar and at periastron passage. Taking into account the toroidal (beyond the light cylinder radius $r_{\rm LC}=c P / 2\pi$)  and dipolar (within light cylinder) nature of the magnetic field, the intensity at the pulsar surface is $B_0\approx 20\ (3\times10^{11} {\rm\, cm} / r_{\rm LC}) (r_{\rm LC} / r_{\rm ns})^3 \approx 1.4\times 10^{12}\,(P/0.1\,\rm s)^2\,$G where $P$ is the pulsar spin period and for a neutron star radius $r_{\rm ns}\approx 10^6\,$cm. This value of $B_0$, for a given $P=0.1\,$s, is standard for rotation-powered gamma-ray pulsars \citep{2013ApJS..208...17A}. 

The total injected power in particles $\dot{E}_p\approx\rm\,10^{35}\, erg\,s^{-1}$ of our best model is also standard for  pulsars detected in gamma rays \citep{2013ApJS..208...17A}. However, combining Eq.~\ref{eq:b1} for the magnetic field with $\dot{E}=\dot{E}_p (1+\sigma)$, we find that our model requires $\sigma\approx 1$. The pulsar spindown power is equally spread  between magnetic and kinetic energy. Such a value is much higher than has been usually assumed in pulsar wind nebulae, starting with the work of \citet{Kennel:1984gu}, though it is not necessarily surprising since pulsar winds are thought to be launched with very high values of $\sigma$ and to convert the magnetic energy to kinetic energy as they propagate (the ``$\sigma$ problem", see e.g. the reviews by \citealt{Kirk:2007wh} and \citealt{Arons:2011aa}). The shock is much closer to the pulsar in LS 5039 ($\approx 3\times10^{11}\,$cm) than in pulsar wind nebulae (0.1 pc in the Crab nebula) so a higher value of $\sigma$ is not problematic {\em per se}. A more worrying issue is that with $\sigma\approx 1$ the assumption of hydrodynamics breaks down. The higher magnetic field at the termination shock means that less of the pulsar wind energy will be transferred to the particles. A full RMHD simulation should be carried out to investigate whether a substantial magnetisation  alleviates some of the difficulties we have encountered in reproducing the observations.

Finally, a pulsar spindown power of $\dot{E}=\rm\,2\times10^{35}\, erg\,s^{-1}$ implies a stellar wind mass loss rate $\dot{M}_w=5\times10^{-9}\rm\,M_\odot\,yr^{-1}$ given $\eta=0.1$ and $v_w=2000\rm\,km\,s^{-1}$. This is at the low end of the range of estimated $\dot{M}_w$, even if we take into account that the spindown power would have to be increased by a factor at least 3 to account for the  1-30 MeV luminosity of $\approx 6\times 10^{35}\,(D/2.5\rm\,kpc)^2\, erg\,s^{-1}$ (assuming we could reproduce the peculiar spectral shape). The mass loss rate estimated from H$\alpha$ measurements give values in the range $2-75\times10^{-8}\rm\,M_\odot\,yr^{-1}$ \citep{2004ApJ...600..927M,Casares:2005gg,2011MNRAS.411.1293S}. However, wind clumping is known to bias this estimator, leading to mass loss rates that can be overestimated by a factor $\approx 20$ for O6.5 stars like LS 5039 \citep{2006ApJ...637.1025F}. Indeed, the lack of signatures from X-ray thermal emission or absorption in the stellar wind favours the lower end of the range of estimated $\dot{M}_w$ \citep{2011MNRAS.411..193S,2011ApJ...743....7Z}. Alternatively, $\eta$ may be smaller than the value we assumed.

\section{Conclusions}
We have developped a post-processing radiative code  to investigate high-energy non-thermal emission based on relativistic hydrodynamical simulations. Our code includes synchrotron emission, anisotropic inverse Compton emission, the opacity due to pair production at VHE, and takes into account relativistic effects using the velocity field from the simulation. The particle energy distribution is evolved according to the adiabatic losses derived from the simulation and radiative losses. Our goal was to provide a coherent model of the spectral modulations observed from X-ray to VHE gamma rays in the gamma-ray binary LS 5039.

(i) The simulation shows a complex shock structure even when orbital motion is neglected. The pulsar flow is fully confined by a bow shock and a back shock. The presence of the back shock induces a reflected shock in the bow shock region. The back shock and reflected shock have a limited impact on the overall emission.

(ii) The VHE emission remains very concentrated towards the apex of the bow shock and strongly absorbed by pair production at superior conjunction. The back shock contribution dominates the VHE flux at superior conjunction but its flux is insufficient to explain the H.E.S.S. detection without emission from the pair cascade. The back shock thus has a very minor influence on the gamma-ray emission from the system. The X-ray emission region is much larger, which will help smoothe out X-ray absorption signatures from the stellar wind \citep{2011MNRAS.411..193S}.

(iii) Doppler boosting plays the major role in modulating the X-ray and VHE emission with orbital phase. Its impact is predominantly set by  the inclination of the system $i$, with double-peaked lightcurves expected at high $i$. We constrain the inclination of LS 5039 to $i\approx 35\degr$.

(iv) There is a tension between the hard VHE spectrum and the level of X-ray emission as they require differing intensities of the magnetic field. This issue is aggravated by the recent COMPTEL detection that, if fully attributed to LS 5039, implies an even stronger synchrotron component (hence higher $B$) and a sharp cutoff between 10 and 100 MeV. These observations cannot be accommodated in our current model. Possible options that may ease the issue include: missing X-ray emission from the simulation box, a more intense magnetic field in the regions where radiative cooling is strong (leading to a denser flow and a more compressed $B$), contributions from the pair cascade triggered by the absorption of VHE gamma rays, a latitude or distance-dependent magnetisation $\sigma$ or wind Lorentz factor $\Gamma_p$.

(v) We attribute the {\em Fermi}-LAT emission component to particles randomized to a Maxwellian distribution at the shock, as shown by simulations of particle acceleration at pair-dominated shocks. This implies  that the Lorentz factor of the wind is $\Gamma_p\approx 5000$. We find that these particles represent the bulk of the power injected in particles, with only 12\% going to the electrons accelerated to a power law. Synchrotron emission from the Maxwellian population produces a weak ($\sim$ 1 mmag) orbital modulation of the optical flux.

(vi) The power-law electrons radiate very efficiently, while the  ``thermal" particles lose energy primarily through adiabatic losses. A modest overall injected power of a few 10$^{35}\rm\,erg\,s^{-1}$ is sufficient to account for the broad band X-ray and TeV emission. For our choice of $\eta=0.1$ this implies a stellar wind mass loss rate of the order of $10^{-8} \rm\,M_\odot\,yr^{-1}$ at the low end of currently estimated values.

(vii) This power, combined with the  magnetic field intensity required by  our best model, implies a pulsar magnetisation $\sigma\approx 1$. Such a high value supports the picture that has pulsar winds launched with high $\sigma$, but fails our assumption of hydrodynamics. Relativistic MHD simulations will be required to further investigate the issue and, perhaps, resolve some of the difficulties encountered in reproducing the observations.

While gamma-ray binaries may be expected to shed light into the processes involved in propagation and termination of pulsar winds, we believe that robust conclusions will require the type of coherent approach linking dynamical and radiative aspects that we have explored here.

\section*{Acknowledgments}
We thank Geoffroy Lesur for his advice and for allowing computations on his private machine. This work was partly supported by the European Community via contract ERC-StG-200911, by the French ``Programme National Hautes Energies'', and by the Centre National d'Etudes Spatiales. AL is supported by the UWM Research Growth Initiative, the NASA ATP program through NASA grant NNX13AH43G, and NSF grant AST-1255469. The RHD simulations have been performed using HPC resources from GENCI- [CINES] (Grant 2013046391) and Texas Advanced Computing Center (TACC) at the University of Texas at Austin (Grant TG-AST-130004).
\bibliographystyle{aa}
\bibliography{../../BIBLIO}
\listofobjects
\end{document}